\documentclass{article}
\usepackage{fancyhdr}
\usepackage[centertags]{amsmath}
\usepackage{amsfonts}
\usepackage{graphics}
\usepackage{graphicx}
\usepackage{wrapfig}
\usepackage{amssymb}
\usepackage{amsthm}
\usepackage{newlfont}
\usepackage{url}
\usepackage{epsfig}
\usepackage{ifthen}
\usepackage{ifpdf}
\usepackage{rotating}
\usepackage{caption}
\usepackage{amsmath}
 \usepackage{subfloat}
\input amssym.def
\input amssym
\input cyracc.def

\newtheorem{lemma}{Lemma}[section]

\newtheorem{te}{Theorem}[section]
\newtheorem{cor}{Corollary}[section]

\newcommand{\be}{\begin{equation}}
\newcommand{\ee}{\end{equation}}
\newcommand{\ba}{\begin{array}}
\newcommand{\ea}{\end{array}}
\newcommand{\bee}{\begin{eqnarray*}}
\newcommand{\eee}{\end{eqnarray*}}
\newcommand{\bea}{\begin{eqnarray}}
\newcommand{\eea}{\end{eqnarray}}

\newcommand{\wmin}{w_{min}}
\newcommand{\wmax}{w_{max}}
\newcommand{\II}{\mathbb{I}}
\newcommand{\ZZ}{\mathbb{Z}}
\newcommand{\LL}{\mathbb{L}}
\newcommand{\GG}{\mathbb{G}}

\begin{document}

\title{
Newton-like method with  diagonal correction for distributed optimization
}


\author{Dragana Bajovi\'c \footnote{Department of Power, Electronics and ´
Communication Engineering, Faculty of Technical Sciences, University of Novi Sad,
Trg Dositeja Obradovi\'ca 6, 21000 Novi Sad, Serbia. email: dbajovic@uns.ac.rs.}
\footnote{Biosense Institute, University of Novi Sad, Ul. Zorana Djindji\'ca 3, 21000 Novi Sad, Serbia.} \and Du\v{s}an Jakoveti\'c \footnotemark[3] \and
Nata\v sa  Kreji\'c
\footnote{Department of Mathematics and Informatics, Faculty of Sciences, University of Novi
  Sad, Trg Dositeja Obradovi\'ca 4, 21000 Novi Sad, Serbia. e-mail:
   \{djakovet@uns.ac.rs, natasak@uns.ac.rs, natasa.krklec@dmi.uns.ac.rs\}. Research supported by the Serbian Ministry of Education,
Science, and Technological Development, Grant no. 174030}
\and
Nata\v{s}a Krklec Jerinki\'c \footnotemark[3]}

\maketitle

\vspace{-1cm}

 \begin{abstract}
 We consider distributed optimization problems where networked nodes cooperatively minimize
 the sum of their locally known convex costs. A popular class of methods to solve these problems
 are the distributed gradient methods, which are attractive due to their inexpensive iterations, but have a drawback
 of slow convergence rates. This motivates the incorporation of second order information in the distributed methods,
 but this task is challenging: although the Hessians which arise in the algorithm design respect the sparsity of the network,
 their inverses are dense, hence rendering distributed implementations difficult. We overcome this challenge and propose a class of distributed Newton-like
 methods, which we refer to as Distributed Quasi Newton (DQN). The DQN family approximates the Hessian inverse
 by: 1) splitting the Hessian into  its diagonal and  off-diagonal part, 2) inverting the diagonal part, and
 3) approximating the inverse of the off-diagonal part through a weighted linear function.
  The approximation is parameterized by the tuning variables which correspond to different splittings of the Hessian
 and by different weightings of the off-diagonal Hessian part. Specific choices of the tuning variables give rise
 to different variants of the proposed general DQN method -- dubbed DQN-0, DQN-1 and DQN-2 --
 which mutually trade-off communication and computational costs for convergence.
 Simulations demonstrate the effectiveness of the proposed DQN methods.

\vspace{3mm}

[2mm]


\end{abstract}

\section{Introduction}

We consider a connected network with $ n $ nodes, each of which has access to a local cost function $ f_i: \mathbb{R}^p \rightarrow \mathbb{R}, \; i=1,\ldots,n. $ The objective for all nodes is to minimize the aggregate cost function  $ f: \mathbb{R}^p \rightarrow \mathbb{R} $, defined by

\be \label{objective}
f(y) = \sum_{i=1}^{n} f_i(y).
\ee

Problems of this form arise in many emerging applications like
 big data analytics, e.g.,~\cite{ScutariBigData},
 distributed inference in sensor networks \cite{RibeiroADMM1, SoummyaEst, SayedConf,
SayedEstimation},
and distributed control, \cite{JoaoMotaMPC}.

Various methods for solving~(\ref{objective}) in a distributed manner are available in the literature. A class of methods based on gradient descent at each node and exchange of information between neighboring nodes is particularly popular, see  \cite{nedic_T-AC, nedic_novo, nedic-gossip, arxivVersion, cdc-submitted, AnnieChen, WotaoYinDisGrad, WotaoYinExtra, ASU_Math_Prog}.
   Assuming that the local costs $f_i$'s are strongly convex and have Lipschitz
  continuous gradients and that a constant step size $\alpha$ is used, these methods converge linearly
  to a solution neighborhood. With such methods, step size~$\alpha$ controls the tradeoff between the convergence speed towards a solution neighborhood and the distance of the limit point from the actual solution, larger $\alpha$ means faster convergence but larger distance from the solution in the limit; see, e.g., \cite{cdc-submitted}, \cite{ribeiro}.
      Distributed first order (gradient) methods allow for a penalty interpretation, where the distributed
    method is interpreted as a (centralized) gradient
   method applied on a carefully constructed penalty reformulation of the original problem~(\ref{objective}); see \cite{cdc-submitted}, \cite{ribeiro} for details.


Given the existence of well developed theory and efficient implementations of higher order methods in centralized optimization in general, there is a clear need to investigate the possibilities of employing higher order methods in distributed optimization as well. More specifically, for additive cost structures~(\ref{objective}) we study here, a further motivation for developing distributed higher order methods comes from their previous success
when applied to similar problems in the context of centralized optimization.
 For example, additive cost~(\ref{objective}) is typical
in machine learning applications where second order methods play an important role, see, e.g.,~\cite{noc1, noc2,noc3}.
Another similar class of problems arise in stochastic optimization, where the objective function is given in the form of mathematical expectation. Again, second order methods are successfully applied in centralized optimization, \cite{schmidt,kk,kkj,ribeirobfgs1,ribeirobfgs2}.

There have been several papers on distributed Newton-type methods.  A distributed second order methods for network utility maximization and network flow optimization are developed in \cite{ErminWei} and~\cite{AccelDualAscent}  but on problem formulations different from~(\ref{objective}).
  Network Newton method \cite{ribeiro} aims at solving (\ref{objective}) and presents a family of distributed (approximate) Newton methods. The class of Netwrok Newton method, refereed to as NN,
   is  extensively analyzed in \cite{ribeiroNNpart1,ribeiroNNpart2}.
  The proposed methods are based on the penalty interpretation, \cite{cdc-submitted,ribeiro},
  of the distributed gradient method in~\cite{nedic_T-AC}, and they approximate
  the Newton step through an $\ell$-th order Taylor approximation of the Hessian inverse, $\ell=0,1,...$
  This approximation gives rise to different variants of methods within the family, dubbed NN-$\ell$. Different choices of
  $\ell$ exhibit inherent tradeoffs between the communication cost and the number of iterations
   until convergence, while NN-$0$, $1$, and $2$ are the most efficient in practice.
   The proposed methods converge linearly to a solution neighborhood,
    exhibit a kind of quadratic convergence phase, and show significantly better simulated performance when compared
     with the standard distributed gradient method in~\cite{nedic_T-AC}.
Reference~\cite{DQM}
       proposes a distributed second order method
       which approximates the (possibly expensive) primal updates with
       the distributed alternating direction of multipliers method in~\cite{DadmmLinear}.
        In~\cite{NewtonRaphsonConsensus},
        the authors propose a distributed Newton Raphson
        method based on the consensus algorithm and separation of time-scales.
         Reference~\cite{ESOM} proposes distributed second order methods
         based on the proximal method of multipliers~(PMM).
         Specifically, the methods approximate the primal variable update step through a second order approximation
          of the NN-type~\cite{ribeiro}. While the NN methods in \cite{ribeiro} converge to a solution neighborhood,
          the methods in \cite{ESOM,DQM,NewtonRaphsonConsensus} converge
            to \emph{exact} solutions.

In this paper, we extend~\cite{ribeiro,ribeiroNNpart1,ribeiroNNpart2} and propose a different family of distributed Newton-like methods for solving~(\ref{objective}).
 We refer to the proposed family as Distributed Quasi
Newton methods~(DQN). The methods are designed to exploit the specific structure of the penalty reformulation~\cite{cdc-submitted,ribeiro}, as is done in~\cite{ribeiro}, but with a different Hessian inverse approximation, for which the idea originates in~\cite{kl}.
Specifically, the  Hessian matrix is  approximated by its block diagonal part, while the remaining part of the Hessian is used to correct the right hand side of the quasi Newton equation. The methods exhibit linear convergence to a solution neighborhood under a set of standard assumptions for the functions $ f_i $ and the network architecture -- each $f_i$ is strongly convex and has Lipschitz continuous gradient, and the underlying network is connected.
 Simulation examples on (strongly
convex) quadratic and logistic losses demonstrate that DQN compares favorably
with NN proposed in~\cite{ribeiro}.

With the DQN family of methods, the approximation of the Newton step is parameterized by
diagonal matrix~$\LL_k$ at each iteration $ k $,  and different choices of $\LL_k$ give rise to different variants of DQN,
which we refer to as DQN-$0$, $1$, and $2$. Different
variants of DQN, based on different matrices $ \LL_k $, tradeoff the number of iterations and
computational cost.
 In particular, setting $\LL_k=0$
 yields the DQN-0 method; a constant, diagonal
 matrix $\LL_k=\LL$ corresponds to DQN-1.
 Finally, $\LL_k$ with DQN-2
 is obtained through
 approximately fitting the Newton
 equation  \eqref{eqn-newton-equation} using a first order Taylor
 approximation.
 The DQN-1 method utilizes the latter,
 DQN-2's weight matrix at the first iteration, and
 then it ``freezes'' it to this constant
 value throughout the iterations; that is,
 it sets $\LL=\LL_0$, where $\LL_0$
 corresponds to the weight matrix of DQN-2 in the initial iteration.

Let us  further specify the main differences between the proposed DQN family and NN methods in \cite{ribeiro} as the NN methods are used as the benchmark in the work presented here.  First,
   the DQN methods introduce
 a different, more general splitting of Hessians with respect to
 NN, parameterized with a scalar~$\theta \geq 0$; in contrast,
  the splitting used in NN corresponds to setting
 $\theta=1$. Second,
 with the proposed variants of DQN-$0$, $1$, and $2$, we utilize \emph{diagonal matrices}
 $\LL_k$, while the NN-$\ell$ methods use in general block-diagonal
 or neighbor-sparse matrices.
 Third, DQN and NN utilize different inverse Hessian approximations;
 the NN's inverse Hessian approximation matrix is symmetric, while
 with DQN it is not symmetric in general. 
   Fourth, while NN approximates the inverse Hessian directly 
   (independently of the Newton equation), DQN  
  actually aims at approximating the Newton equation. 
  Hence, unlike NN, the resulting DQN's inverse Hessian approximation (with DQN-2 in particular) explicitly depends on the gradient
  at the current iterate, as it is the case with many Quasi-Newton methods; see,  e.g.,~\cite{DennisSchnabel}.
  Finally, the analysis
 here is very different from~\cite{ribeiro}, and the  major reason comes
 from the fact that the Hessian approximation with DQN is asymmetric in general.
 This fact also incurs the need for a safeguarding step with DQN in general, as detailed in Section~3.
  We also point out that results presented in~\cite{ribeiro} show that NN methods
  exhibit a quadratic convergence phase. It is likely that similar
  results can be shown for certain variants of DQN methods as well,
  but detailed analysis is left for future work. It may be very challenging to rigorously compare the linear convergence rate factors of DQN and NN methods
 and their respective inverse Hessian approximations. However, we provide in Section~5 both certain analytical insights and extensive numerical experiments to compare the two classes of methods.

As noted, DQN methods
do not converge to the exact solution of~(1), but they converge to a solution neighborhood,
as it is the case with other methods (e.g., distributed gradient descent~\cite{nedic_T-AC} and NN methods~\cite{ribeiro})
which are based on
the penalty interpretation of~(1). Hence,
for very high accuracies, they may not be competitive
with distributed second order methods which
 converge to the exact solution~\cite{DQM,ESOM,NewtonRaphsonConsensus}.
 However, following the framework
 of embedding distributed second
 order methods into PMM -- developed in~\cite{ESOM}, we
 apply here the DQN Newton direction approximations to the PMM methods; we refer to the resulting
 methods as PMM-DQN-$\ell$, $\ell=0,1,2$.
 Simulation examples on strongly convex quadratic costs demonstrate
 that the PMM-DQN methods compare favorably with the methods in~\cite{DQM,ESOM,NewtonRaphsonConsensus}.
 Therefore, with respect to the existing literature and in particular with respect to \cite{ribeiro,ESOM}, this paper broadens the possibilities for distributed approximations of relevant Newton directions, and hence offers alternative distributed second order methods, which exhibit competitive performance on the considered simulation examples.   Analytical studies of PMM-DQN are left for future work.

This paper is organized as follows. In Section 2 we give the problem statement and some preliminaries needed for the definition of the method and convergence analysis. Section 3 contains the description of the proposed class of Newton-like methods and convergence results. Specific choices of the diagonal matrix that specifies the method completely are presented in Section 4. Some simulation results are presented in Section 5 while Section~6 discusses extensions of embedding DQN in the PMM framework. Finally, conclusions are drawn in Section 7, while Appendix provides some auxiliary derivations.

\section{Preliminaries}
Let us first give some preliminaries about  the problem (\ref{objective}),  its penalty interpretation in~\cite{cdc-submitted,ribeiro}, as well as the decentralized gradient descent algorithm in~\cite{nedic_T-AC} that will be used later on.

The following assumption on the $f_i$'s is imposed.

\noindent {\bf Assumption A1.} The functions $f_{i} : \mathbb{R}^p \rightarrow \mathbb{R}, \; i=1,\ldots,n$ are twice continuously differentiable, and there exist constants $0<\mu\leq L<\infty$ such that for every $x \in \mathbb{R}^p$
$$\mu I \preceq \nabla^{2} f_{i}(x) \preceq L I.$$
 Here, $I$ denotes the $p \times p$ identity matrix, and notation~$M  \preceq N$ means that the matrix $N-M$ is positive semi-definite.

This assumption implies that the functions $ f_i , \; i=1,\ldots,n $ are strongly convex with modulus $ \mu > 0, $
\be \label{A11}
f_{i}(z) \geq f_{i}(y) + \nabla f_{i}(y)^T (z-y) + \frac{\mu}{2} \|z-y\|^2, \; y,z \in \mathbb{R}^p,
\ee
and the gradients are Lipschitz continuous with the constant $ L $ i.e.
\be
\label{A12}
\|\nabla f_i(y) - \nabla f_i(z)\| \leq L \|y-z\|, \; y,z \in \mathbb{R}^p, \; i=1,\ldots, n.
\ee

 Assume that the network of nodes is an undirected network $ {\cal G} = ({\cal V},{\cal E}), $ where $ {\cal V} $ is the set of nodes and $ {\cal E} $ is the set of all edges,  i.e., all pairs $\{i,j\} $ of nodes which can exchange information through a communication link.

 \noindent {\bf Assumption A2.}
 The network  $ {\cal G} = ({\cal V},{\cal E}) $ is connected, undirected and simple (no self-loops nor multiple links).

 Let us denote by $ O_i $ the set of nodes that are connected with the node $ i $ (open neighborhood of node $i$) and let $\bar{O}_{i}=O_{i}\bigcup \{i\}$ (closed neighborhood of node $i$).  We associate with $ {\cal G} $ a symmetric, (doubly) stochastic $n \times n$ matrix $ W. $ The elements of $ W $ are all nonnegative and rows (and columns) sum up to one. More precisely,  we assume the following.

 \noindent{\bf Assumption A3.} The matrix $ W = W^T \in \mathbb{R}^{n \times n} $ is stochastic
 with elements $ w_{ij} $ such that
 $$ w_{ij} > 0 \mbox{ if } \{i,j\} \in {\cal E}, \;  w_{ij} = 0 \mbox{ if }  \{i,j\} \notin {\cal E},\,i \neq j, \mbox{ and }  w_{ii} = 1 - \sum_{j \in O_i} w_{ij} $$
 and there are constants $\wmin$ and $\wmax$ such that for $i=1,\ldots,n$
 $$0 < \wmin \leq w_{ii} \leq  \wmax <1.$$

 Denote  by $ \lambda_1 \geq \ldots \geq \lambda_n $ the eigenvalues of $ W. $ Then it can be easily seen that $ \lambda_1 = 1.  $
  Furthermore, the null space of $ I - W $ is spanned by $ e := (1,\ldots,1).  $

 Following \cite{cdc-submitted}, \cite{ribeiro},  the  auxiliary function $ \Phi : \mathbb{R}^{n p} \rightarrow \mathbb{R}, $ and the corresponding penalty reformulation of (\ref{objective}) is introduced as follows. Let $ x=(x_1,\ldots,x_n) \in \mathbb{R}^{np} $ with $ x_i \in {\mathbb{R}}^p, $ and denote by  $ \ZZ \in \mathbb{R}^{np \times np} $ the matrix obtained as the Kronecker product of $ W $ and the identity $ I \in \mathbf{R}^{p\times p}, \ZZ = W \otimes I. $

 The corresponding  penalty reformulation of (\ref{objective}) is given by
  \be \label{reformulation1}
 \min_{x \in {\mathbb{R}^{np}}} \Phi(x) := \alpha \sum_{i=1}^{n} f_i(x_i) + \frac{1}{2} x^T({\mathbb {I}} - \ZZ)x.
 \ee
  Applying the standard gradient method to (\ref{reformulation1}) with the unit step size we get
  \be
 \label{dgd}
 x^{k+1} = x^k - \nabla \,\Phi(x^k),\,k=0,1,...,
 \ee
which, denoting the $i$-th $p \times 1$ block of $x^k$ by $x_i^k,$  and after rearranging terms, yields
 the Decentralized Gradient Descent (DGD) method \cite{nedic_T-AC}
 \be \label{dgd}
 x_i^{k+1} = \sum_{j \in \bar{O}_i} w_{ij} \, x_j^k - \alpha \nabla f_i(x_i^k), \; i=1,\ldots, n.
 \ee
 Clearly, the penalty parameter $ \alpha $ influences the relation between (\ref{reformulation1}) and (\ref{objective}) -- a smaller $ \alpha $ means better agreement between the problems but also implies smaller steps in (\ref{dgd}) and thus makes the convergence slower. It can be shown, \cite{ribeiro} that if $ \tilde{y} \in \mathbb{R}^p $ is the solution of (\ref{objective}) and $ x^{*}:=(\bar{y}_1,\ldots,\bar{y}_n) \in \mathbb{R}^{n p} $ is the solution of (\ref{reformulation1}) then, for all $i$,
 \be \label{ocena1}
 \|\bar{y}_i - \tilde{y}\| = {\cal O}(\alpha).
 \ee
 The convergence of (\ref{dgd}) towards $x^*$ is linear, i.e., the following estimate
   holds \cite{WotaoYinDisGrad,cdc-submitted},
 \be \label{ocena2}
 \Phi(x^k) - \Phi(x^{*}) \leq (1-\xi)^k (\Phi(x^0) - \Phi(x^*)),
 \ee
 where $ \xi \in (0,1) $ is a constant depending on $ \Phi, \alpha $ and $ W. $

The matrix and vector norms that will be frequently used in the sequel are defined here.
Let $ \|a\|_2 $ denote the Euclidean norm of vector~$a$ of arbitrary dimension.
Next, $\|A\|_2$ denotes the spectral norm of matrix~$A$ of arbitrary dimension.
%
 Further, for a matrix ${\mathbb M} \in \mathbb{R}^{np\times np}$ with blocks $M_{ij} \in  \mathbb{R}^{p\times p}$, we
 will also use
 the following block norm:
$$\|{\mathbb M}\|:= \max_{j =1,\ldots,n} \sum_{i=1}^{n} \|M_{ij}\|_{2},$$
where, as noted, $\|M_{ij}\|_{2}$ is the spectral norm of $M_{ij}$. For a vector $x \in \mathbb{R}^{np}$ with blocks $x_{i} \in  \mathbb{R}^{p}$,
 the following block norm is also used:
$$\|x\|:=\sum_{i=1}^{n} \|x_{i}\|_{2}.$$

\section{Distributed Quasi Newton method}
In this section we introduce a class of Quasi Newton methods for solving (\ref{reformulation1}).  The general Distributed Quasi Newton (DQN) method is proposed in Subsection 3.1. The method is characterized by a generic diagonal matrix $ \LL_k. $ Global linear convergence rate for the class is established in Subsection 3.2, while  local linear convergence rate with the full step size $\varepsilon=1$ is analyzed in Subsection 3.3.
Specific variants DQN-$0$, $1$, and $2$, which correspond to the specific choices of $\LL_k$, are studied in Section 4.
 As we will see, algorithm DQN has certain tuning parameters, including the  step size~$\varepsilon$. Discussion on the tuning parameters choice is
 relegated to Section 4.

\subsection{The proposed general DQN method}

The problem  we consider from now on is~(\ref{reformulation1}), where we recall $ \ZZ = W \otimes I $  and $W$ satisfies assumption A3.

The problem under consideration  has a specific structure as the Hessian is  sparse if the underlying network is sparse. However its inverse is dense. Furthermore, the matrix inversion (i.e. linear system solving) is not suitable for decentralized computation. One possibility of exploiting the structure of $ \nabla^2 \Phi(x) $ in a distributed environment is presented in \cite{ribeiro} where the Newton step is approximated through a  number of inner iterations. We  present here a different possibility.  Namely, we keep the diagonal part of $ \nabla^2 \Phi(x) $ as the Hessian approximation but at the same time use the non-diagonal part of $ \nabla^2 \Phi(x) $ to correct the right hand side vector in the (Quasi)-Newton equation. Let us define the splitting
$$W_{d}=diag(W) \quad \mbox{and} \quad W_u=W-W_{d},$$
and $\ZZ=\ZZ_{d}+\ZZ_{u}$ with
$$ \ZZ_{d}=W_{d} \otimes I=diag(\ZZ) \quad \mbox{and} \quad  \ZZ_{u}=W_{u} \otimes I.$$
 Here, $diag(W)$ denotes the diagonal matrix
 with the same diagonal as the matrix $W$.
 Hence, matrix $\ZZ_d$ is
 a $np \times np$ diagonal matrix whose
 $i$-th $p \times p$ block is the scalar matrix $w_{ii}I,$
 $\ZZ_u$ is a $np \times np$ block (symmetric) matrix such that
 $(i,j)$-th $p \times p$ off-diagonal blocks are again scalar matrices  $w_{ij}I$, while the diagonal blocks are
 all equal to zero.

Clearly, the gradient is $$ \nabla \Phi(x) = \alpha \nabla F(x) + (\II-\ZZ)x,  $$ where
$\II$ denotes the $np \times np$ identity matrix and
$$ F(x) = \sum_{i=1}^{n} f_i(x_i), \quad \nabla F(x) = (\nabla f_1(x_1),\ldots,\nabla f_n(x_n)), $$
while the Hessian is
$$ \nabla^2 \Phi(x) = \alpha \nabla^{2} F(x) + \II-\ZZ$$
where $\nabla^{2} F(x)$ is the block diagonal matrix with the $i$th diagonal block $\nabla^2 f_i(x_i)$.

The general Distributed Quasi Newton, DQN, algorithm is presented below. Denote by
$k$ the iteration counter, $k=0,1,...$, and let
$ x^k = (x_1^k,...,x_n^k)\in {\mathbb R}^{n p}$ be the estimate of
$x^*$ at iteration~$k$. Consider the following splitting of the Hessian
\be \label{hes} \nabla^2 \Phi(x^k)= {\mathbb A}_k  - \GG, \ee
with
\be \label{Ak}
{\mathbb A}_k =\alpha \nabla^{2} F(x^k) + (1+\theta)(\II-\ZZ_{d})
\ee
and
$$\GG=\ZZ_{u}+\theta (\II-\ZZ_{d})$$
for some $\theta \geq 0$. Hence,
$\GG$
is a~$np \times np$ block (symmetric) matrix
whose $i$-th $p \times p$
diagonal block equals $g_{ii}\,I$, with $g_{ii}:=\theta\,(1-w_{ii})$,
while the $(i,j)$-th
$p \times p$
off-diagonal block equals $g_{ij}\,I$, with $g_{ij}:=w_{ij}$. One can easily see that the splitting above recovers the splitting for NN methods \cite{ribeiro}  taking $ \theta = 1. $ We keep $ \theta $ unspecified for now and later on, we will demonstrate numerically that taking $ \theta=0 $ can be beneficial.
 Also, notice that $ {\mathbb A}_k $ is  block diagonal with the $i$th diagonal block
$$A^{k}_i=\alpha \nabla^2 f_i(x^{k}_i)+(1+\theta)(1-w_{ii})I.$$

Let $\LL_{k} \in \mathbb{R}^{np \times np}$ be a diagonal matrix composed of diagonal $ p \times p $ matrices $ \Lambda_i^k, \; i=1,\ldots,n.  $
 In this paper, we adopt
 the following approximation of the Newton direction
  $s^k_N = -({\mathbb A}_k -\mathbb{G})^{-1} \nabla \Phi(x^k)$:
\be \label{QN1}
s^k = -(\II - \LL_k \GG) {\mathbb A}^{-1}_k \nabla \Phi(x^k).
\ee
The motivation for this approximation  comes from~\cite{kl} and the following reasoning.
 Keep the Hessian approximation on the left hand side of the Newton equation
 $({\mathbb A}_k -\mathbb{G}) s^k_N = -\nabla \Phi(x^k)$
  diagonal,
 and correct the right hand side through the off-diagonal
Hessian part. In more detail, the Newton equation can be equivalently written as:
\begin{equation}
\mathbb{A}_k \,s_N^k  = - \nabla \Phi(x^k) + \mathbb{G}\,s^k_N,
\end{equation}
where the off-diagonal part $\widetilde{s}^k:=\mathbb{G}\,s^k_N$ is moved
to the right hand side.
If we pretended for a moment to know
the value of $\widetilde{s}^k$,
then the Newton direction is obtained as:
\begin{equation}
 \label{eqn-cover-letter-1}
s^k_N  = - \left( \mathbb{A}_k \right)^{-1} \left( \nabla \Phi(x^k) - \widetilde{s}^k \right).
\end{equation}
This form is suitable
for distributed implementation due to the need to invert only the block
diagonal matrix $\mathbb{A}_k$.
However, $\widetilde{s}^k$ is clearly
not known, and hence we approximate it.
 To this end, note that,
 assuming that $\mathbb{G}\nabla \Phi(x^k)$
 is a vector with all the entries being non-zero,
 without loss
 of generality,
 $\widetilde{s}^k$ can be written as follows:
 \begin{equation}
 \label{eqn-cover-letter-2}
 \widetilde{s}^k = \mathbb{L}_k\,\mathbb{G}\,\nabla \Phi(x^k),
 \end{equation}
  where $\mathbb{L}_k$ is a \emph{diagonal matrix}.
  Therefore, estimating $\widetilde{s}^k $
   translates into estimating the diagonal matrix~$\mathbb{L}_k$,
   assuming that $\mathbb{G}\,\nabla \Phi(x^k)$ is known.
    We follow~\cite{kl}
    and consider simple
    approximations of the ``true'' $\mathbb{L}_k$.
     E.g., we will consider $\mathbb{L}_k = 0$
      which discards the off-diagonal Hessian part.
      Also, as we will see ahead, we adopt a  Taylor approximation method for estimating $\mathbb{L}_k$.
    Now, substituting  (\ref{eqn-cover-letter-2}) into
    (\ref{eqn-cover-letter-1}), we obtain
    the following Newton direction approximation:
    \begin{equation}
    \label{eqn-cover-letter-3}
    s_0^k  = - \left( \mathbb{A}_k \right)^{-1} \left( \mathbb{I} - \mathbb{L}_k \,\mathbb{G} \right)
     \nabla \Phi(x^k).
    \end{equation}
   Finally, we adopt (\ref{QN1}) as the definite form of the
   Newton direction approximation,
   i.e., we permute the matrices $\left( \mathbb{I} - \mathbb{L}_k \,\mathbb{G} \right)$
    and $\left( \mathbb{A}_k \right)^{-1}$.
    The reason is that both
    $s^k$ and $s_0^k$ have the same
    inner product with $\nabla \Phi(x^k)$,
    and hence they have the same descent
    properties (for example, Theorem~{3.2}
     ahead holds unchanged for $s^k$ and $s_0^k$),
     while a careful inspection shows that $s^k$ allows for
     a cheaper (in terms of
     communications per iteration) distributed implementation.

     The reasoning
     above justifies restriction to \emph{diagonal} $\mathbb{L}_k$'s, i.e.,
     in view of~(\ref{eqn-cover-letter-2}), adopting diagonal $\mathbb{L}_k$'s
     does not \emph{in principle, i.e., in structure} sacrifice the quality
     of the Newton direction approximation, while it is
     computationally cheap.

Then, following  a typical Quasi-Newton scheme, the next iteration is defined by
\be \label{QN2}
x^{k+1} = x^k + \varepsilon s^k,
\ee for some step size $ \varepsilon. $

Clearly, the choice of $ \LL_k $ is crucial in the approximation of $ (\nabla^2 \Phi(x^k))^{-1}. $
The following general algorithm assumes only that $ \LL_k $ is diagonal and bounded. Specific choices of $ \LL_k $ will be discussed in Section 4.
 Actually, all the proposed variants DQN-$0$, $1$, and $2$ utilize \emph{diagonal} matrices $\LL_k$.
 Parameter $ \theta $ affects splitting (\ref{hes}) and the search direction (\ref{QN1}). For this moment we are assuming only that $ \theta $ is nonnegative and fixed initially, while  further details are presented later on.

In summary, the proposed distributed algorithm for solving (\ref{reformulation1}) is given below.

 \noindent{\bf Algorithm 1: DQN in vector format}\\
 Given $ x^0 \in \mathbb{R}^{np},  \varepsilon, \rho > 0. $
 Set $ k = 0. $
 \begin{itemize}
 \item[Step 1.] Chose a diagonal matrix $\LL_{k} \in \mathbb{R}^{np \times np} $ such that $$\|\LL_{k}\| \leq \rho.$$
 \item[Step 2.]
Set
$$  s^k = - (\II - \LL_k \GG) {\mathbb A}_k^{-1} \nabla \Phi(x^k). $$
 \item[Step 3.] Set
 $$ x^{k+1} = x^k + \varepsilon s^k, \; k = k+1. $$
 \end{itemize}

For the sake of clarity, the proposed algorithm,
from the perspective of each node~$i$ in the network, is presented in Algorithm~2.


\noindent{\bf Algorithm 2: DQN -- distributed implementation}\\
\noindent{At each node $i$, require $\rho,\varepsilon>0. $
\begin{itemize}
\item[1] Initialization: Each node $i$ sets $k=0$ and $x_i^0 \in {\mathbb R}^p$.
\item[2] Each node $i$ transmits $x_i^k$ to all its neighbors
    $j \in O_i$ and receives $x_j^k$ from all $j \in O_i$.
    \item[3]Each node $i$ calculates
    \[
    d_i^k = \left( A_i^k\right)^{-1} \left[ \, \alpha\, \nabla f_i(x_i^k)
    + \sum_{j \in O_i} w_{ij} \left( x_i^k - x_j^k\right)\,\right].
    \]
    \item[4] Each node~$i$ transmits $d_i^k$ to all its neighbors
    $j \in O_i$ and receives $d_j^k$ from all $j \in O_i$.
    \item[5] Each node $i$ chooses a diagonal $p \times p$ matrix $\Lambda_i^k$, such that
    $ \|\Lambda_i^k\|_2 \leq \rho. $
    \item[6] Each node $i$ calculates:
    \[
    s_i^k  = - d_i^k + \Lambda_i^k \sum_{j \in \bar{O}_i} G_{ij} \,d_j^k.
    \]
    \item[7] Each node $i$ updates its solution estimate as:
    \[
    x_i^{k+1} = x_i^k + \varepsilon\,s_i^k.
    \]
    \item[8] Set $k=k+1$ and go to step 3.
\end{itemize}

Calculation of $\Lambda_i^k$ in step~6 will be specified in the next section, and, for certain algorithm variants,
will involve an additional inter-neighbor communication of a $p$-dimensional vector.
 Likewise, choices of tuning parameters $\varepsilon, \rho, \theta$ are discussed
 throughout the remaining of this section and Section 4.

%

\noindent{\bf Remark}.
It is useful to compare~(\ref{QN1}) with the direction adopted in NN methods.
Setting $\theta=1$ and $\LL_k=0$ recovers NN-$0$,
$\theta=1$ and $\LL_k=-{\mathbb A}_k^{-1}$ recovers NN-$1,$ while
NN-$2$ can not be recovered in a similar fashion. Hence, DQN in a sense
generalizes NN-$0$ and NN-$1$. An approximation $ \mathbb{L}_k = \mathbb{L}, $ that will be considered later on, does not recover any of NN methods.

Observe that the approximation matrix
$(\II - \LL_k \GG) \mathbb{A}_k^{-1}$ is not symmetric in general. This fact  induces the need for
a safeguarding step in Algorithm~1, more precisely the elements of $ \mathbb{L}_k $ are uniformly bounded
as stated in Step~1 (Algorithm~1),  $ \|\mathbb{L}\|  \leq \rho, $ and the resulting method requires an analysis different from~\cite{ribeiro}.

\subsection{Global linear convergence rate}

In this subsection, the  global linear convergence rate of Algorithm  DQN is established.
The convergence analysis consists of two parts. First,  we  demonstrate that $ s^k $ is a descent direction. Then we determine a suitable interval for the step size $ \varepsilon $ that ensures linear convergence of the iterative sequence.

The following Gershgorin type theorem for block matrices is needed for the first part of convergence analysis.

\begin{te} \cite{FV} \label{Varga}
For any $ \mathbb{C} \in \mathbb{R}^{np \times np} $ partitioned into blocks $ C_{ij} $ of size $p\times p$, each eigenvalue $ \mu $ of $ \mathbb{C} $ satisfies
\be \label{FV1}  \frac{1}{\|(C_{ii} - \mu I)^{-1}\|_2} \leq \sum_{i \neq j} \|C_{ij}\|_2 \ee for at least one $ i \in \{1,\ldots,n\}. $
\end{te}
Using the above theorem we can prove the following lower bound for all eigenvalues of a symmetric block matrix.
\begin{cor} \label{cor1}
Let $ \mathbb{C} \in \mathbb{R}^{np \times np}  $ be a symmetric matrix partitioned into blocks $ C_{ij} $ of size $p\times p$. Then each eigenvalue $ \mu $ of $ \mathbb{C} $ satisfies
$$ \mu \geq \min_{i=1,\ldots,n} \left\{ \lambda_{\min}(C_{ii}) - \sum_{j \neq i} \|C_{ij}\|_2\right\}, $$ where $ \lambda_{\min}(C_{ii}) $ is the smallest eigenvalue of $ C_{ii}. $
\end{cor}

{\em Proof. } Given that $ \mathbb{C} $ is symmetric, all its eigenvalues are real. Also, $ C_{ii} $ is symmetric and has only real eigenvalues. Now, fix one eigenvalue $ \mu $ of the matrix  $ \mathbb{C}. $ By Theorem \ref{Varga}, there exists $ i \in  \{1,\ldots,n\}, $ such that (\ref{FV1}) holds. Next, we have
$$ \|(C_{ii} - \mu I)^{-1}\|_2 =  \frac{1}{\min_{j=1,\ldots,p} |\lambda_j(C_{ii}) - \mu|}, $$ where $ \lambda_j(C_{ii}) $ is the $j$-th eigenvalue of $ C_{ii}. $
Thus
$$ \min_{j=1,\ldots,p} |\lambda_j(C_{ii}) - \mu| \leq \sum_{j \neq i} \|C_{ij}\|_2. $$ We have just concluded that, for any eigenvalue $ \mu $ of $ \mathbb{C} $ there exists $ i \in \{1,\ldots,n\} $ and $ j \in \{1,\ldots,p\} $ such that $\mu $ lies in the interval
$$ [\lambda_j(C_{ii}) - \sum_{i \neq l} \|C_{il}\|_2, \lambda_j(C_{ii}) + \sum_{i \neq l} \|C_{il}\|_2]. $$ Hence, for each $ \mu $ for which (\ref{FV1}) holds for some fixed $ i, $ we have
$$ \mu \geq \lambda_{\min}(C_{ii}) - \sum_{l \neq i} \|C_{il}\|_2 $$ and the statement follows. $ \Box $

We are now ready to prove that the search direction (\ref{QN1}) is descent.
\begin{te} \label{th31} Suppose that A1-A3 hold.  Let
   \begin{equation}
   \label{eqn-rho}
   0 \leq \rho \leq \frac{\alpha \mu +(1+\theta)(1-\wmax)}{(1-\wmin)(1+ \theta)}\left(\frac{1}{\alpha L+(1+\theta)(1-\wmin)}-\delta\right)
   \end{equation}
   for some $\delta \in (0,1/(\alpha L+(1+\theta)(1-\wmin)))$.
Then $ s^k $ defined by (\ref{QN1}) is a descent direction and satisfies
 $$ \nabla^T \Phi(x^k) s^k \leq - \delta \|\nabla \Phi(x^k)\|_2^2. $$
\end{te}

{\em Proof.}
Let us first show that   $s^k$ is  descent search direction.
As $$  \nabla^{T} \Phi(x^k) s^{k} = - \nabla^T \Phi(x^k)  (\II - \LL_k \GG) {\mathbb A}_k^{-1} \nabla \Phi(x^k), $$ $ s^k $ is descent if  $ v^T(\II - \LL_k \GG) {\mathbb A}_k^{-1} v > 0 $ for arbitrary $ v \in \mathbb{R}^{np \times np}. $
Given that $ (\II - \LL_k \GG) {\mathbb A}_k^{-1}$   is not symmetric in general, we know the above is true if and only if the matrix $$ \mathbb{C}^k = \frac{1}{2}( (\II - \LL_k \GG){\mathbb A}_k^{-1}+{\mathbb A}_k^{-1}(\II - \GG \LL_k)) $$ is positive definite. $\mathbb{C}^k $
is symmetric and thus it should be positive definite if all of its eigenvalues are positive.
The matrix  $\mathbb{C}^{k}$ is partitioned in the blocks
$$C^{k}_{ii}=(A^{k}_{i})^{-1}-\frac{1}{2}\theta (1-w_{ii})(\Lambda^{k}_{i}(A^{k}_{i})^{-1}+(A^{k}_{i})^{-1}\Lambda^{k}_{i}), \quad i=1,\ldots,n,$$
$$C^{k}_{ij}=-\frac{1}{2} w_{ij} (\Lambda^{k}_{i}(A^{k}_{j})^{-1}+(A^{k}_{i})^{-1}\Lambda^{k}_{j}),\quad i \neq j.$$
Corollary \ref{cor1}  implies that
$$\lambda_{min}(\mathbb{C}^k)\geq \min_{i=1,\ldots,n} (\lambda_{min}(C^{k}_{ii})-\sum_{j\neq i}\|C^{k}_{ij}\|_{2})$$
The definition of $\mathbb{A}_{k}$ implies
\be \label{novo}
(\alpha \mu +(1+\theta) (1-w_{ii}))I \preceq A^{k}_{i} \preceq (\alpha L +(1+\theta) (1-w_{ii}))I
\ee
so,
$$(\alpha \mu +(1+\theta) (1-\wmax))I \preceq A^{k}_{i} \preceq (\alpha L +(1+\theta) (1-\wmin))I$$
and therefore, for every $i=1,\ldots,n$
$$\|(A^{k}_{i})^{-1}\|_{2} \leq \frac{1}{\alpha \mu +(1+\theta) (1-\wmax)}.$$
Moreover, it follows
$$\lambda_{min}(C^{k}_{ii}) \geq \frac{1}{\alpha L +(1+\theta) (1-\wmin)}-\frac{\theta (1-w_{ii})  \rho}{\alpha \mu +(1+\theta) (1-\wmax)}$$ and
$$\|C^{k}_{ij}\|_{2} \leq \frac{w_{ij} \rho}{\alpha \mu +(1+\theta) (1-\wmax)}.$$
Now,
\begin{eqnarray}
\lambda_{min}(C^k) & \geq  & \min_{i=1,\ldots,n} (\frac{1}{\alpha L +(1+\theta) (1-\wmin)}-\frac{\theta (1-w_{ii})  \rho}{\alpha \mu +(1+\theta) (1-\wmax)} \nonumber \\
& - & \sum_{j \in O_{i}}w_{ij} \frac{ \rho}{\alpha \mu +(1+\theta) (1-\wmax)}) \nonumber \\
& = &  \min_{i=1,\ldots,n} (\frac{1}{\alpha L +(1+\theta) (1-\wmin)} -\frac{\rho (1-w_{ii})(1+ \theta)}{\alpha \mu +(1+\theta) (1-\wmax)})\nonumber \\
& \geq  &  \frac{1}{\alpha L +(1+\theta) (1-\wmin)} -\rho \frac{(1-\wmin)(1+ \theta)}{\alpha \mu +(1+\theta) (1-\wmax)} \nonumber\\
& \geq  &   \delta. \label{G1}
\end{eqnarray}
Since $\delta>0$ we conclude that $\mathbb{C}^{k}$ is positive definite. Moreover, $v^{T} \mathbb{C}^{k} v = v^T(\II - \LL_k \GG)\mathbb{A}_k^{-1} v, $  for any $ v \in \mathbb{R}^{np \times np} $ and
\begin{eqnarray}
\nabla^{T} \Phi(x^k) s^{k} & = & - \nabla^T \Phi(x^k)  (\II - \LL_k \GG) {\mathbb A}_k^{-1} \nabla \Phi(x^k) \nonumber \\
& = & - \nabla^T \Phi(x^k) C^{k} \nabla \Phi(x^k) \nonumber \\
& \leq  &  - \delta \|\nabla \Phi(x^k)\|_2^2.\nonumber \\
\end{eqnarray}
 $ \Box $

The next lemma corresponds to the standard property of descent direction methods that establish the relationship between the search vector and the gradient.

\begin{lemma} \label{lema2}
 Suppose that A1-A3 hold. Then
 $$ \|s^k\|_2 \leq \beta \|\nabla \Phi(x^k)\|_2, $$
 where
 \be \label{beta} \beta=\frac{1+\rho (1+\theta) (1-\wmin)}{\alpha \mu +(1+\theta) (1-\wmax)}.\ee
 \end{lemma}

{\em Proof.} For matrix $\mathbb{A}_{k}$, there holds that:
\be \label{akm} \|\mathbb{A}_{k}^{-1}\|_2  \leq \frac{1}{\alpha \mu +(1+\theta) (1-\wmax)}.\ee
This can be shown similarly as the upper bound on
$\|({A}^{k}_i)^{-1}\|_2$ below~(19).
%
%
Furthermore,
\be \label{ggb} \|\GG\|_2 \leq (1+\theta) (1-\wmin).\ee
This is true
as
\begin{eqnarray*}
\|\mathbb{G}\|_{2} &=& \|\mathbb{Z}_{u}+\theta (\mathbb{I}-\mathbb{Z}_{d})\|_{2}
\leq
 \|\mathbb{Z}_{u}\|_{2}+\theta \| \mathbb{I}-\mathbb{Z}_{d}\|_{2} \\
 &\leq&
 \| \mathbb{I}-\mathbb{Z}_{d}\|_{2} +\theta \| \mathbb{I}-\mathbb{Z}_{d}\|_{2} \leq
(1+\theta) (1-w_{min}),
\end{eqnarray*}
where we used
$\mathbb{Z}_u=\mathbb{Z}-\mathbb{Z}_d \preceq
\mathbb{I}-\mathbb{Z}_d \preceq  (1-w_{min})\mathbb{I}$.
  Therefore, we have:
 \begin{eqnarray*}
 \|s^k\|_2 & \leq  & \|(\II-\LL_k \GG){\mathbb A}_k^{-1}\|_2 \|\nabla \Phi(x^k)\|_2 \\
 & \leq & (1+\|\LL_k\|_2 \| \GG\|_2) \|{\mathbb A}_k^{-1}\|_2 \|\nabla \Phi(x^k)\|_2 \\
 &\leq & \frac{1+\rho (1+\theta) (1-\wmin)}{\alpha \mu +(1+\theta) (1-\wmax)}  \|\nabla \Phi(x^k)\|_2.
 \end{eqnarray*} $ \Box$

 Let us now show that there exists a step size $ \varepsilon > 0 $ such that the sequence $ \{x^k\} $ generated by Algorithm DQN converges to the solution of (\ref{reformulation1}). Notice that (\ref{reformulation1}) has a unique solution, say $ x^*. $ Assumption A1 implies that $ \nabla \Phi(x) $ is  Lipschitz continuous as well, i.e., with $ \tilde{L}:=\alpha L+2(1-\wmin), $ there holds
 \be \label{Ltilda}
 \|\nabla \Phi(x) - \nabla \Phi(y)\|_2 \leq \tilde{L} \|x - y\|_2, \; x,y \in \mathbb{R}^{np}.
 \ee
 Furthermore,  \be \label{mu}
 \frac{\tilde{\mu}}{2}\|x - x^*\|_2^2 \leq \Phi(x) - \Phi(x^*) \leq \frac{1}{\tilde{\mu}} \|\nabla \Phi(x)\|_2^2
 \ee
 for   $\tilde{\mu}=\alpha \mu $ and all $ x \in \mathbb{R}^{np}. $ The main convergence statement is given below.

 \begin{te} \label{the} Assume that the conditions of Theorem \ref{th31} are satisfied. Define
 \begin{equation}
 \label{eqn-varepsion-thm-3-3}
  \varepsilon = \frac{\delta}{\beta^{2} \tilde{L}}
  \end{equation}
 with $\beta$ given by (\ref{beta}).
 Then Algorithm DQN generates a sequence $ \{x^k\} $ such that
 $$ \lim_{k \to \infty} x^k = x^* $$ and the convergence is at least linear with
 $$ \Phi(x^{k+1}) - \Phi(x^*) \leq \left(1 - \frac{ \delta^2 \tilde{\mu}}{2 \tilde{L}\beta^2}\right) \left( \Phi(x^k) - \Phi(x^*)\right), \quad k=0,1,\ldots. $$
 \end{te}

 {\em Proof. }
 The Mean Value Theorem, Lipschitz property of $ \nabla \Phi $, Theorem \ref{th31}  and Lemma \ref{lema2} yield
 \begin{eqnarray}
 \Phi(x^{k+1}) - \Phi(x^*) & = & \Phi(x^k + \varepsilon s^k) - \Phi(x^*) \nonumber \\
 & = & \Phi(x^k) + \int_0^1 \nabla^T \Phi(x^k + t \varepsilon s^k) \varepsilon s^k dt - \Phi(x^*) \pm \varepsilon \nabla^T \Phi(x^k) s^k \nonumber \\
 & \leq & \Phi(x^k) - \Phi(x^*) + \varepsilon \int_{0}^1 \|\nabla^T \Phi(x^k + t \varepsilon s^k)  - \nabla^T \Phi(x^k)\|_2 \|s^k\|_2 dt \nonumber \\
 &  + & \varepsilon \nabla^T \Phi(x^k) s^k \nonumber \\
 & \leq & \Phi(x^k) - \Phi(x^*) + \varepsilon \int_0^1 \tilde{L} t \varepsilon \|s^k\|_2^2 dt + \varepsilon \nabla^T \Phi(x^k) s^k \nonumber \\
 & = & \Phi(x^k) - \Phi(x^*) + \frac{1}{2} \varepsilon^2 \tilde{L} \|s^k\|_2^2 +  \varepsilon \nabla^T \Phi(x^k) s^k \nonumber \\
 & \leq & \Phi(x^k) - \Phi(x^*) + \beta^{2} \frac{\tilde{L}}{2}  \varepsilon^2 \| \nabla \Phi(x^k)\|_2^2 -  \varepsilon \delta \|\nabla \Phi(x^k)\|_2^2 \nonumber \\
 & = & \Phi(x^k) - \Phi(x^*) + \left( \frac{\beta^{2} \tilde{L}}{2} \varepsilon^2 - \varepsilon \delta\right)\|\nabla^2 \Phi(x^k)\|_2^2. \label{fix}
\end{eqnarray}
Define
$$ \phi(\varepsilon) =  \frac{\beta^{2} \tilde{L}}{2} \varepsilon^2 - \varepsilon \delta. $$ Then $ \phi(0) = 0, \; \phi'(\varepsilon) = \tilde{L} \varepsilon \beta^{2}-\delta$ and $ \phi''(\varepsilon) > 0. $ Thus, the minimizer of $ \phi $ is $ \varepsilon^* = \delta/(\beta^{2} \tilde{L}) $ and
\be \label{fie}
 \phi(\varepsilon^*) = - \frac{\delta^{2}}{2 \beta^{2} \tilde{L}}.
\ee
Now, (\ref{fix}) and (\ref{fie}) give
$$  \Phi(x^{k+1}) - \Phi(x^*) \leq  \Phi(x^k) - \Phi(x^*) -  \frac{\delta^{2}}{2 \beta^{2} \tilde{L}} \|\nabla \Phi(x^k)\|_2^2. $$ From (\ref{mu}), we also have
$$ \Phi(x^k) - \Phi(x^*) \leq \frac{1}{\tilde{\mu}} \|\nabla \Phi(x^k)\|_2^2 $$
and
$$ -  \frac{\delta^{2}}{2 \beta^{2} \tilde{L}} \|\nabla \Phi(x^k)\|_2^2 \leq - \left( \Phi(x^k) - \Phi(x^*) \right)  \frac{\delta^{2}\tilde{\mu}}{2 \beta^{2} \tilde{L}}, $$ so
$$  \Phi(x^{k+1}) - \Phi(x^*) \leq \left(1 -  \frac{\delta^{2} \tilde{\mu}}{2 \beta^{2} \tilde{L}}\right) \left(\Phi(x^k) - \Phi(x^*) \right). $$
Given that $\tilde{\mu} =\alpha \mu \leq \alpha L<\tilde{L}$, we have $\tilde{\mu} /\tilde{L}<1$.
Moreover,
\begin{eqnarray*} \delta & < & \frac{1}{\alpha L+(1+\theta)(1-\wmin)} \leq \frac{1}{\alpha \mu +(1+\theta)(1-\wmax)} \\
& \leq & \frac{1+\rho (1+\theta) (1-\wmin)}{\alpha \mu +(1+\theta) (1-\wmax)}=\beta
\end{eqnarray*}
and
$$ \xi:=1 -  \frac{\delta^{2}\tilde{\mu}}{2 \beta^{2} \tilde{L}}  \in (0,1) .$$
We conclude with
$$  \Phi(x^{k+1}) - \Phi(x^*) \leq \xi \left( \Phi(x^k) - \Phi(x^*) \right) $$ and
$$ \lim_{k \to \infty} \Phi(x^k) = \Phi(x^*). $$ As $ \Phi \in C^2(\mathbb{R}^n), $ the above limit also implies $$ \lim_{k \to \infty} x^k = x^*. $$
$\Box$

 The proof of the above theorem clearly shows that for any $\varepsilon \in (0,\delta/(\beta^{2} \tilde{L})]$  algorithm DQN converges. However, taking $\varepsilon$ as large as possible implies larger steps and thus faster convergence.

\subsection{Local linear convergence }

We have proved global linear convergence for the specific step length $\varepsilon$ given in Theorem \ref{the}.
However, local linear convergence can be obtained for the full step size, using the theory developed for Inexact Newton methods \cite{DES}. The step $ s^k $ can be considered as an Inexact Newton step and we are able to estimate the residual in Newton equation as follows.

\begin{te} \label{th33} Suppose that A1-A3 hold. Let $ x^k $ be such that $ \nabla \Phi(x^k) \neq 0.  $ Assume that  $ s^k $ is generated in Step 2 of Algorithm~1 with
$$ 0 \leq \rho < \frac{ \alpha \mu}{(1+\theta) (1-\wmin)\left(\alpha L+2(1+\theta)(1-\wmin)\right)}. $$
 Then there exists  $t \in (0,1) $ such that
$$ \|\nabla \Phi(x^k) + \nabla^2 \Phi(x^k) s^k \| < t \|\nabla \Phi(x^k)\|. $$
\end{te}

{\em Proof.} First, notice that  the interval for $\rho$ is well defined. The definition of the search direction (\ref{QN1}) and the splitting of the Hessian (\ref{hes}) yield
$$\nabla \Phi(x^k) + \nabla^2 \Phi(x^k) s^k= (  \GG {\mathbb A}_k^{-1}+{\mathbb A}_k \LL_{k} \GG {\mathbb A}_k^{-1}-\GG \LL_{k} \GG {\mathbb A}_k^{-1}   ) \nabla \Phi(x^k):=\mathbb{Q}_{k} \nabla \Phi(x^k) .$$
Therefore,
$$\|\mathbb{Q}_{k}\| \leq  \| \GG {\mathbb A}_k^{-1}\|+\| \GG {\mathbb A}_k^{-1}\| \| \LL_{k}\| \|{\mathbb A}_k\| + \| \GG {\mathbb A}_k^{-1}\| \|\LL_{k}\| \| \GG \|. $$
Moreover,
\begin{eqnarray*}
 \|\GG {\mathbb A}_k^{-1}\|&=& \max_{j} (\theta (1-w_{jj})\|(A_{j}^{k})^{-1}\|+\sum_{i \in O_{j}} w_{ij}\|(A_{j}^{k})^{-1} \|) \\
&\leq & \max_{j} \frac{\theta (1-w_{jj})+1-w_{jj}}{\alpha \mu + (1+\theta)(1-w_{jj})}.\\
\end{eqnarray*}
Recalling (\ref{novo}) and the fact that the expression above is decreasing with respect to $ w_{jj}, $ we get
\begin{equation} \label{eqn-spec-rad-GAinv}
 \|\GG {\mathbb A}_k^{-1}\| \leq   \frac{(1+\theta) (1-\wmin)}{\alpha \mu + (1+\theta)(1-\wmin)} =: \gamma,
\end{equation}
and there holds $\|A_{k}\| \leq \alpha L+(1+\theta)(1-\wmin)$.
Furthermore,~(\ref{akm}), (\ref{ggb}) and  $\|\LL_{k}\| \leq \rho$ imply
\begin{eqnarray}
\|\mathbb{Q}_{k}\| &\leq & \gamma+ \gamma\rho (1+\theta)(1-\wmin) +\gamma \rho (\alpha L+(1+\theta)(1-\wmin)) \nonumber \\
&=& \gamma + \rho \gamma (\alpha L+2(1+\theta)(1-\wmin)) \label{26novo} \\
&<& \gamma+1-\gamma=1. \nonumber
\end{eqnarray}
 Thus,  the statements is true with
 \begin{equation}
 \label{eqn-to-comment}
 t=\gamma + \rho \gamma (\alpha L+2(1+\theta)(1-\wmin)).
 \end{equation}
$\Box$

Theorem~{3.4} introduces an upper bound on the safeguard parameter $ \rho$
different than the one considered in Theorem~{3.2}.
The relation between the two bounds depends on
the choice of $\delta$ in Theorem~{3.2}.
Taking a sufficiently small $\delta$ in Theorem~{3.2}, we obtain that
 $\rho$ in Theorem~{3.2} is larger.
 However, taking $\delta < \frac{1}{\alpha L + (1+\theta)(1-w_{min})}$
 sufficiently close to $\frac{1}{\alpha L + (1+\theta)(1-w_{min})}$,
 $\rho$ in Theorem~{3.4} eventually becomes larger.

One way to interpret the relation between Theorems~3.2 and 3.3
on one hand, and Theorem~3.4 on the other hand, as far as $\rho$
is concerned, is as follows.
Taking a very small $\delta$,
Theorem~{3.3} allows
for a quite large $\rho$
 but on the other hand
 it significantly decreases the admissible step size $\varepsilon$.
 At the same time,
 Theorem~{3.4} corresponds in a sense
 to an opposite
 situation where $\varepsilon$
  is allowed to be quite large (in fact, equal to one),
  while $\rho$ is quite restricted.
 Therefore, the two results exploit the
 allowed ``degrees of freedom'' in a different way.

%

For the  sake of completeness we list here the conditions for local convergence of Inexact Newton methods.

\begin{te} \label{thDES} \cite{DES} Assume that A1 holds and that $ s^k $ satisfies the inequality
 $$ \|\nabla \Phi(x^k) + \nabla^2 \Phi(x^k) s^k \| < t \|\nabla \Phi(x^k)\|, \quad k=0,1,\ldots $$ for some $ t<1. $
 Furthermore, assume that $ x^{k+1} = x^k + s^k, \; k=0,1,\ldots. $ Then there exists $ \eta > 0 $ such that for all $ \|x^0 - x^*\| \leq \eta, $ the sequence $ \{x^k\} $ converges to $ x^*. $ The convergence is linear,
 $$ \|x^{k+1} - x^*\|_* \leq t \|x^k - x^*\|_*, k=0,1,\ldots, $$ where
 $ \|y\|_* = \|\nabla^2 \Phi(x^*) y \|. $
 \end{te}

 The two previous theorems imply the following Corollary.

 \begin{cor} \label{cor32} Assume that the conditions of Theorem \ref{th33} hold. Then there exists $\eta >0$ such that for every $x^{0}$  satisfying $\|x^{0}-x^{*}\| \leq \eta, $ the sequence $ \{x^k\} $ generated by Algorithm DQN and $ \varepsilon = 1$
 converges linearly to $ x^* $ and
 $$ \|\nabla^2 \Phi(x^*) (x^{k+1} - x^{*})\| \leq t \|\nabla^2 \Phi(x^*) (x^k - x^*)\|, \quad k=0,1,\ldots $$
 holds with $t \in (0,1)$.
 \end{cor}

For (strongly convex) quadratic functions $f_{i}$, $i=1,\ldots,n $ we can also claim  global linear convergence as follows.

\begin{te}  \label{th36} Assume that all loss functions $ f_i $ are strongly convex quadratic and that the conditions of Theorem \ref{th33} are satisfied. Let $ \{x^k\} $ be a sequence generated by Algorithm DQN with  $ \varepsilon = 1.  $ Then $ \lim_{k \to \infty} x^k = x^* $ and
$$ \|x^{k+1} - x^*\|_* \leq t\,\|x^k - x^*\|_*, \quad k=0,1,\ldots $$  for $ t $ defined in Theorem \ref{th33}.
\end{te}

{\em Proof. }  Given that the penalty term in~(\ref{reformulation1}) is convex quadratic,  if all local cost functions $f_i$ are strongly convex quadratic, then  the objective function $\Phi$ is also strongly convex quadratic,
i.e., it can be written as
\be \Phi(x) = \frac{1}{2} (x-x^*)^T \mathbb{B} (x-x^*), \label{quadratic} \ee
for some fixed, symmetric positive definite matrix $ \mathbb{B} \in \mathbb{R}^{np \times np}. $
Recall that $x^*$ is the global minimizer of $\Phi$. Then
$$ \nabla \Phi(x) = \mathbb{B} (x-x^*) \mbox{ and } \nabla^2 \Phi(x) = \mathbb{B}. $$ Starting from
$$ s^k = - (\nabla^2 \Phi(x^k))^{-1} \nabla \Phi(x^k) + e^k, $$ we get
\begin{eqnarray*}
 \|\nabla^2 \Phi(x^k) e^k\| & = & \|\nabla^2 \Phi(x^k) \left(s^k + (\nabla^2 \Phi(x^k))^{-1} \nabla \Phi(x^k)\right)\| \\
  &  = & \|\nabla \Phi(x^k) + \nabla^2 \Phi(x^k) s^k \| < t \|\nabla \Phi(x^k)\| \end{eqnarray*} by Theorem \ref{th33}.
  Next,
 \begin{eqnarray*}   x^{k+1} & = & x^k + s^k = x^k - (\nabla^2 \Phi(x^k))^{-1} \nabla \Phi(x^k) + e^k \\
 & = & x^k - \mathbb{B}^{-1} \nabla \Phi(x^k) + e^k, \end{eqnarray*}
 and
 $$ x^{k+1} - x^* = x^k - x^* - \mathbb{B}^{-1} \nabla \Phi(x^k) + e^k. $$
 Therefore,
 $$ \mathbb{B} (x^{k+1} - x^*) = \mathbb{B}(x^k - x^*) - \nabla \Phi(x^k)  + \mathbb{B}e^k. $$
Now,
$$  \|\mathbb{B} e^k\| = \|\nabla^2 \Phi(x^k) e^k \| < t \| \nabla \Phi(x^k)\| = t \|\mathbb{B}(x^k - x^*)\|, $$ and
$$ \|\mathbb{B}(x^{k+1} - x^*)\| = \|\mathbb{B}e^k\| \leq t \|\mathbb{B}(x^k - x^*)\|. $$ $ \Box $

\section{Variants of the general DQN} 

Let us now discuss the possible alternatives for the choice of $ \LL_k. $ Subsection 4.1  presents three different variants of the general DQN algorithm which
mutually differ in the choice of matrix~$\LL_k$. We refer to the three choices as
DQN-$0$, DQN-$1$, and DQN-$2$. All results established in Section 3 hold for
these three alternatives. Subsection 4.1 also provides local linear convergence
rates for DQN-$2$ without safeguarding. Subsection 4.2
 gives a discussion on the algorithms' tuning parameters,
 as well as on how the required global knowledge by all nodes
 can be acquired in a distributed way.

\subsection{Algorithms DQN-$0$, $1$, and~$2$}

The analysis presented so far implies only that the diagonal matrix $ \LL_k $ has to be bounded. Let us now look closer at different possibilities for defining $ \LL_k, $ keeping the restrictions stated in Theorem \ref{th31} and \ref{th33}.

\textbf{DQN-0}.  We first present the method DQN-$0$ which sets $\LL_k=0$. Clearly, for DQN-$0$, Theorems
3.2 and 3.4 hold, and thus we get linear convergence with the proper choice of $ \varepsilon, $ and local linear convergence with $ \varepsilon =1. $
The approximation of the Hessian inverse in this case equals $ \mathbb{A}_k^{-1} $, i.e., the Hessian is approximated by its block diagonal part only.
 The method DQN-$0$ corresponds to Algorithm~1
 with only steps 1-4 and 7-9 executed, with $\Lambda_i^k=0$ in step~7.
 Clearly, choice $\LL_k=0$ is the cheapest possibility among the choices of $\LL_k$ if we consider the computational cost per iteration~$k$. The same holds for communication cost per~$k$, as each node
needs to transmit only $x_i^k$ per each iteration, i.e., one $p$-dimensional vector per node, per iteration is communicated.
 We note that DQN-$0$ resembles NN-$0$, but the difference in general is that
 DQN-$0$ uses a different splitting, parameterized with $\theta \geq 0$; actually,
 NN-$0$ represents the special case with~$\theta=1$.

\textbf{DQN-1}. Algorithm DQN-$1$ corresponds to setting $ \LL_k = \LL, k=0,1,\ldots,$ where $\LL$ is a constant
\emph{diagonal} matrix. Assuming that $ \LL $ is chosen such that $ \|\LL\| \leq \rho, $ with $ \rho $ specified in Theorem \ref{th31},  global linear convergence for a proper step size $\varepsilon$ and local linear convergence for the full step size $\varepsilon=1$ again hold. Algorithm DQN-$1$ is given by Algorithm~1, where each node
utilizes a constant, diagonal matrix~$\Lambda_i$. There are several possible ways of
choosing the $\Lambda_i$'s. In this paper,
we focus on the following choice. In the first iteration $k=0$, each node $i$
sets matrix $\Lambda_i^0$ through algorithm DQN-$2$, stated in the sequel, and then
it keeps
the same matrix $\Lambda_i^0$ throughout the whole algorithm.
 The computational cost per iteration of DQN-$1$ is higher than the cost of DQN-$0$.
At each iteration, each node $ i $ needs to compute the corresponding inverse of $i$-th block of $ \mathbb{A}_k $ and then to multiply it by the constant diagonal matrix $\Lambda_i$.
 Regarding the communication cost, each node transmits two $p$-dimensional vectors per iteration -- $x_i^k$ and $d_i^k$
  (except in the first iteration $k=0$ when it also transmits
  an additional vector $u_i^0$; see ahead Algorithm~3).
  Although the focus of this paper is on the diagonal $\LL_k$'s, we
  remark that setting $\theta=1$ and $\LL_k=-{\mathbb A}_k^{-1}$ recovers the NN-1 method.


\textbf{DQN-2}. Algorithm DQN-$2$ corresponds to an iteration-varying,
diagonal matrix $\LL_k$.
 Ideally, one would like to choose matrix $ \LL_k $ such that search direction $ s^k $ resembles the Newton step as much as possible, with the restriction that $ \LL_k $ is diagonal. The Newton direction $ s^k_{N} $ satisfies the equation
\be \label{eqn-newton-equation} \nabla^2 \Phi(x^k) s_N^k + \nabla \Phi(x^k) = 0. \ee
We seek $ \LL_k $ such that it makes residual $ M(\LL_k) $ small, where
$ M(\LL_k) $ is defined as follows:
\be \label{lambda1}
M(\LL_k) =  \|\nabla^2 \Phi(x^k) s^k + \nabla \Phi(x^k)\| .
\ee
 Notice that
 \begin{eqnarray*}
 M(\LL_k) & = & \|\nabla^2 \Phi(x^k) s^k + \nabla \Phi(x^k)\| \\
 & = & \|-\nabla^2 \Phi(x^k)(I - \LL_k \GG){\mathbb A}_k^{-1} \nabla \Phi(x^k) + \nabla \Phi(x^k)\| \\
 & = & \| -\nabla^2 \Phi(x^k){\mathbb A}_k^{-1} \nabla \Phi(x^k)+\nabla^2 \Phi(x^k)\LL_k \GG {\mathbb A}_k^{-1} \nabla \Phi(x^k) + \nabla \Phi(x^k)\| \\
 &=&  \| -(A_{k}-\GG){\mathbb A}_k^{-1} \nabla \Phi(x^k)+\nabla^2 \Phi(x^k)\LL_k \GG {\mathbb A}_k^{-1} \nabla \Phi(x^k) + \nabla \Phi(x^k)\| \\
 &=&  \| \GG {\mathbb A}_k^{-1} \nabla \Phi(x^k)+\nabla^2 \Phi(x^k)\LL_k \GG {\mathbb A}_k^{-1} \nabla \Phi(x^k) \|.
 \end{eqnarray*}
 Therefore,
 \be \label{u} \nabla^2 \Phi(x^k) s^k + \nabla \Phi(x^k)= u^k + \nabla^2 \Phi(x^k) \LL_k u^k,\ee
 where $$ u^k = \GG {\mathbb A}_k^{-1} \nabla \Phi(x^k) . $$
The minimizer of $ M(\LL_k) $ is clearly achieved if $ \LL_k $ satisfies the equation
 \be \label{lambdak}
 \LL_k u^k = - (\nabla^2 \Phi(x^k))^{-1} u^k,
 \ee
 but (\ref{lambdak}) involves the inverse Hessian. Thus we  approximate $ (\nabla^2 \Phi(x^k))^{-1} $ by the Taylor expansion as follows. 
Clearly,
 \be \label{Bk} (\nabla^2 \Phi(x^k))^{-1} = (\alpha \nabla^2 F(x^k)+\II - \ZZ)^{-1} =  (\II - \mathbb{V}_k)^{-1}, \quad V_k=\ZZ - \alpha \nabla^2 F(x^k). \ee
 Assume that $ \alpha < (1+\lambda_n)/L, $ with $ \lambda_n $  being the smallest eigenvalue of $ W. $
 Then
 \[
 \mathbb{V}_k \succeq (\lambda_n - \alpha\,L)\,\II \succ - \II.
\]
Similarly,
\[
\mathbb{V}_k \preceq (1-\alpha\,\mu)\,\II \prec I.
\]
Hence,
$$ \rho(\mathbb{V}_k) \leq \|\mathbb{V}_k\|_2 <1.$$
  Therefore,  $ \II - \mathbb{V}_k $ is nonsingular,
 $$ (\II-\mathbb{V}_k)^{-1} = \II + \mathbb{V}_k +\sum_{i=2}^{\infty} \mathbb{V}_{k}^i, $$
 and the approximation
 \be \label{Bkapprox}
 (\nabla^2 \Phi(x^k))^{-1}= (\II-\mathbb{V}_k)^{-1} \approx  \II+\mathbb{V}_k
 \ee
  is well defined.  So, we can take $ \LL_k $ which satisfies the following equation
 \be \label{lambda1}
 \LL_k u^k = - (\II + \mathbb{V}_k)u^k.
 \ee
 Obviously, $ \LL_k $ can be computed in a distributed manner.
 We refer to the method which corresponds to this choice of $\LL_k$ as DQN-$2$.
 The algorithm is given by Algorithm~2 where step~6, the choice of $\LL_k=diag(\Lambda_1,...,\Lambda_n),$
  involves the  steps presented below in Algorithm~3. Denote by~$u_i^k$ the $i$-th $p \times 1$
  block of $ u^k $ -- the block which corresponds to node~$i.$

\noindent{\bf Algorithm 3: Choosing $\mathbb{L}_k$ with DQN-2}
\begin{itemize}
    \item[6.1] Each node~$i$ calculates
    \[
    u_i^k = \sum_{j \in \bar{O}_i}G_{ij}\,d_j^k.
    \]
    \item[6.2] Each node~$i$ transmits $u_i^k$ to all its neighbors $j \in O_i$
     and receives $u_j^k$ from all $j \in O_i$.
    \item[6.3] Each node $i$ calculates ${\Lambda}_i^k$ -- the solution to the following
    system of equations (where the only unknown is the $p \times p$
     diagonal matrix $\Lambda_{i}^k$):
    \[
    \Lambda_i^k \,u_i^k = - \left[\,(1+w_{ii})I-\alpha\,\nabla^2 f_i(x_i^k)\,\right]
    \,u_i^k - \sum_{j \in O_i}w_{ij}\,u_j^k.
    \]
    \item[6.4] Each node~$i$ projects each diagonal entry of $ {\Lambda}_i^k$ onto the interval~$[-\rho,\,\rho]$.

\end{itemize}

Note that step~6 with algorithm DQN-$2$ requires an additional $p$-dimensional communication
per each node, per each~$k$ (the communication of the~$u_i^k$'s.) Hence, overall, with
algorithm DQN-$2$ each node transmits three $p$-dimensional vectors per $k$ --
 $x_i^k$, $d_i^k$, and $u_i^k$.

We next show that algorithm DQN-$2$ exhibits local linear convergence even when
safeguarding (Step~6.4 in Algorithm~3) is not used.

\begin{te} \label{th37} Suppose that A1-A3 hold and  let $ x^k $ be an arbitrary point such that $ \nabla \Phi(x^k) \neq 0. $
Assume that \be \label{uslovi}\alpha < \min \left\{ \frac{1+\lambda_n}{L},\,\frac{\wmin}{2L}, \,\frac{2\mu}{L^2} \right\},\ee
and $ s^k $ is generated by (\ref{QN1})  and Algorithm 2, Steps 6.1 -6.3.  Then there exists  $t \in (0,1) $ such that
$$ \|\nabla \Phi(x^k) + \nabla^2 \Phi(x^k) s^k \| < t \|\nabla \Phi(x^k)\|. $$
\end{te}

{\em Proof.} Using (\ref{u}) and  (\ref{lambda1})  we obtain
\begin{eqnarray}
 \|\nabla^2 \Phi(x^k) s^k + \nabla \Phi(x^k)\|& = & \|u^k + \nabla^2 \Phi(x^k) \LL_k u^k\| \nonumber \\
 & = & \|u^k - \nabla^2 \Phi(x^k) (\II + V_k)u^k\| \nonumber \\
 & = & \|(\II-\nabla^2 \Phi(x^k) (\II + \ZZ - \alpha \nabla^2 F(x^k)))u^k\| \nonumber \\
 &=& \|\mathbb{P}^{k} u^k\|, \label{4z}
 \end{eqnarray}
where
\begin{eqnarray*}
 \mathbb{P}^{k} &=& \II - \nabla^2 \Phi(x^k)(\II + \ZZ - \alpha \nabla^2 F(x^k)) \\
& = & \II - \left(\II + \alpha \nabla^2 F(x^k) - \ZZ\right) \left(\II + \ZZ - \alpha \nabla^2 F(x^k)\right) \\
& = &  \ZZ^2 -  \alpha (\ZZ \nabla^2 F(x^k) +\nabla^2 F(x^k) \ZZ)+ \left(\alpha \nabla^2 F(x^k)\right)^2
\end{eqnarray*}
Since $\|\nabla^{2} f_{i} (x_{i})\| \leq L,$ there follows  $\| \nabla^2 F(x^k)\| \leq L$ and the previous
 equality  implies
\be \label{oblik1}
\|\mathbb{P}^{k}\| \leq  \|\ZZ^2 -  \alpha (\ZZ \nabla^2 F(x^k) +\nabla^2 F(x^k) \ZZ)\| + \alpha^2 L^2:=\|U^{k}\|+ \alpha^2 L^2.
\ee
Now, $$U_{ij}^{k}=\sum_{k=1}^{n}w_{ik}w_{kj}I-\alpha(w_{ij}\nabla^2 f_j(x_j^k)+w_{ij}\nabla^2 f_i(x_i^k)).$$
Furthermore,  the assumption $\alpha < \wmin/(2L)$ implies
$$\sum_{k=1}^{n}w_{ik}w_{kj}\geq w_{ii}w_{ij}\geq w_{ij} \wmin \geq w_{ij} 2 \alpha L \geq w_{ij} 2 \alpha \mu.$$
Moreover, $\nabla^2 f_j(x_j^k) \succeq \mu I$ and
$$\|U_{ij}^{k}\|_{2} \leq \sum_{k=1}^{n}w_{ik}w_{kj}-2 \alpha \mu w_{ij}.$$
Therefore,
\begin{eqnarray*}
 \|U^{k}\| &\leq & \max_{j=1,\ldots,n} \sum_{i=1}^{n} (\sum_{k=1}^{n}w_{ik}w_{kj}-2 \alpha \mu w_{ij}) \\
& = & \max_{j=1,\ldots,n} \sum_{k=1}^{n} w_{kj} \sum_{i=1}^{n} w_{ik}-2 \alpha \mu \sum_{i=1}^{n} w_{ij} \\
& = &  1-2 \alpha \mu.
\end{eqnarray*}
So,
\be \label{3z} \|\mathbb{P}^{k}\| \leq h(\alpha), \ee
where $h(\alpha)=1-2 \alpha \mu +\alpha^{2}L^{2}$. This function is convex and nonnegative since $ \mu \leq L$ and therefore
$$\min_{\alpha} h(\alpha)=h\left(\frac{\mu}{L^2}\right)=1-\frac{\mu^2}{L^2}>0.$$
Moreover, $h(0)=h\left(\frac{2 \mu}{L^2}\right)=1$ and we conclude that for all $\alpha \in (0, 2 \mu /L^2)$ there holds  $h(\alpha) \in (0,1)$.
As
$$ u^k = \GG {\mathbb A}_k^{-1}\nabla \Phi(x^k), $$ we have
\be \label{2z}  \|u^k\| \leq \|\GG {\mathbb A}_k^{-1}\| \|\nabla \Phi(x^k)\|. \ee
Now,
\begin{eqnarray*}
 \|\GG {\mathbb A}_k^{-1}\|&=& \max_{j} (\theta (1-w_{jj})\|(A_{j}^{k})^{-1}\|+\sum_{i \in O_{j}} w_{ij}\|(A_{j}^{k})^{-1} \|)\\
&\leq & \max_{j} \frac{\theta (1-w_{jj})+1-w_{jj}}{\alpha \mu + (1+\theta)(1-w_{jj})}\\
&=&  \frac{(1+\theta) (1-\wmin)}{\alpha \mu + (1+\theta)(1-\wmin)} < 1.
\end{eqnarray*}
Therefore,
\be \label{1z}  \|u^k\| < \|\nabla \Phi(x^k)\|. \ee
Putting together (\ref{4z})-(\ref{1z}), for $\theta \geq 0 $ and $\alpha $ satisfying (\ref{uslovi}) we obtain
$$\| \nabla \Phi(x^k) + \nabla^2 \Phi(x^k) s^k\|\leq  h(\alpha) \|u^{k}\|< h(\alpha) \|\nabla \Phi(x^k)\|,$$
i.e. the statement holds with
\begin{equation}
\label{eqn-t-corollary}
t=h(\alpha) =1-2 \alpha \mu +\alpha^{2}L^{2}
\end{equation}
 $\Box$

Applying Theorem \ref{thDES} once again, we get the local linear convergence as stated in the following corollary.

\begin{cor} \label{cor41}
Assume that the conditions of Theorem \ref{th37} hold. Then there exists $\eta$ such that for every $x^{0}$  satisfying $\|x^{0}-x^{*}\| \leq \eta $ the sequence $ \{x^k\}, $ generated by DQN-2 method with Steps 6.1-6.3 of Algorithm~3 and $ \varepsilon = 1,$
 converges linearly to $ x^* $ and
 \begin{equation}
\label{eqn-ineq-cor}
\|\nabla^2 \Phi(x^*) (x^{k+1} - x^{*})\| \leq t \|\nabla^2 \Phi(x^*) (x^k - x^*)\|, \quad k=0,1,\ldots
\end{equation}
 holds with $t$ given by (\ref{eqn-t-corollary}).
 \end{cor}

We remark that, for strongly convex quadratic~$f_i$'s,
the result analogous to Theorem \ref{th36} holds in the sense of global linear convergence, i.e.,
inequality~(\ref{eqn-ineq-cor}) holds for all~$k$ and arbitrary initial point~$x^0$.

\noindent{\bf Remark}.
 ***An interesting future research direction is
to adapt and analyze convergence of the DQN methods in asynchronous environments, 
 as it
  has been already numerically
  studied recently in~\cite{EMR}. Therein, it is shown that
  the studied second order methods still converge in an asynchronous
  setting, though with a lower convergence speed.

\subsection{Discussion on the tuning parameters}

Let us  now comment on the choice of the
involved parameters -- matrix $W$ and scalars $\alpha, \rho, \varepsilon,\theta$, and $\delta$. We first consider
the general DQN method in Algorithm~1, i.e.,
our comments apply to all DQN-$\ell$
 variants, $\ell=0,1,$ and $2$.

Matrix~$W$ only needs to satisfy that: 1) the underlying support network is connected and
 2) all diagonal entries $w_{ii}$ lie between $w_{min}$ and $w_{max}$,
where $0<w_{min} \leq w_{max}<1$. Regarding the latter condition,
it is standard and rather mild; it is only required for~(\ref{ocena1})
to hold, i.e., to ensure that solving~(\ref{reformulation1})
gives an approximate solution to the desired problem~(\ref{objective}).
Regarding the second condition, it can be easily fulfilled through simple
weight assignments, e.g., through the
Metropolis weights choice; see, e.g.,~\cite{BoydFusion}.

We now discuss the choice of the
parameters $\alpha,\rho,\theta,\varepsilon, $ and $\delta$. First, $ \alpha $ defines the penalty reformulation~(4),
 and therefore, it determines the asymptotic error that the algorithm achieves.
  The smaller $\alpha$, the smaller the limiting (saturation)
   error of the algorithm is, but the slower the convergence rate is.
   Thus, the parameter should be set a priori
   according to a given target accuracy; see also~\cite{WotaoYinDisGrad}.
   A practical guidance, as considered in~\cite{NNDDtsp},
   is to set $\alpha=1/(\mathcal{K}\,L)$,
   where~$L$ is the Lipschitz
   gradient constant as in the paper, and
   $\mathcal{K}=10$-$100$.
 %
 %
  %
  Next, parameter $ \theta \geq 0$ determines the splitting of the Hessian.
  It can simply be taken as $\theta=0$, and a justification
  for this choice
  can be found in Section~5.
  Next, the role of $\delta$
   is mainly theoretical.
    Namely, Theorems~{3.2} and~{3.3}
     consider \emph{generic}
        choices of $\mathbb{L}_k$'s, and they are
        worst-case type results.
         Therein, $\delta$ essentially
         trades off the guaranteed
         worst-case (global linear) convergence factor with
         the size of admissible range of the $\mathbb{L}_k$'s
          (size of the maximal allowed $\rho$).
         As per~(\ref{eqn-rho}),
         a reasonable choice to balance
          the two effects is
         $\delta=\frac{1}{2 \left(\alpha L+(1+\theta)w_{min}\right)}$.
         Having set $\delta$,
         the remaining two parameters,
         $\varepsilon$ and $\rho$, can be set according to (\ref{eqn-varepsion-thm-3-3})
          and~(\ref{eqn-rho}), respectively.
         As noted, the described
         choice of the triple
         $(\delta,\rho,\epsilon)$
         is a consequence of the worst case,
         conservative analysis with Theorems~{3.2} and~{3.3}
          (which still have a theoretical significance, though).
         In practice,
         we recommend setting $\delta=0$,
         $\varepsilon=1$, and
         $\rho$ as the upper bound in~(\ref{eqn-rho}) with $\delta=0$.

\textbf{Discussion on distributed implementation}. The algorithm's tuning parameters need to  be set beforehand in a distributed way.
 Regarding weight matrix~$W$, each node $i$ needs to store beforehand
 the weights $w_{ii}$ and $w_{ij}$, $j \in O_i$, for all its neighbors.
 The weights can be set according to the Metropolis rule, e.g., \cite{BoydFusion},
 where each node~$i$ needs to know only the degrees of
 its immediate neighbors. Such weight choice, as noted before,
 satisfies the imposed assumptions.

In order
to set the scalar tuning parameters $\alpha,$ $\theta$,
$\varepsilon$, and $\rho$, each node~$i$ needs
to know beforehand global quantities $w_{min}$, $w_{max}$,
$\mu$ and $L$. Each of these parameters represent either a maximum or a
minimum of nodes' local quantities. For example,
$w_{max}$ is the maximum of the $w_{ii}$'s over $i=1,...,n$,
where node $i$ holds quantity~$w_{ii}$. Hence, each node can obtain
$w_{max}$ by running a distributed algorithm for maximum
computation beforehand; for example, nodes can
utilize the algorithm in~\cite{JohanssonMaximum}.

\section{Simulations}
\label{section-simulations}
This section shows numerical performance of the proposed methods on two examples, namely the
strongly convex quadratic cost functions
and  the logistic loss functions.

\textbf{Simulation setup}. Two simulation scenarios with different types of nodes'cost functions $f_i$'s: 1) strongly convex quadratic costs and 2)
 logistic (convex) loss functions are considered. Very similar scenarios have been considered in~\cite{ribeiro,ribeiroNNpart1,ribeiroNNpart2}.
 With the quadratic costs scenario, $f_i: \,{\mathbb R}^p \rightarrow \mathbb R$ is given by
 $$f_i(x) = \frac{1}{2}(x-a_i)^\top B_i (x-a_i),$$ where $B_i \in {\mathbb R}^{p \times p}$ is a positive definite (symmetric matrix), and $a_i \in {\mathbb R}^p$ is a vector. Matrices $B_i$, $i=1,...,n$ are generated mutually
 independently, and so are the vectors $a_i$'s; also, $B_i$'s are generated independently from the $a_i$'s.
 Each matrix~$B_i$ is generated as follows. First, we generate a matrix $\widehat{B}_i$
  whose entries are drawn mutually independently from the standard normal distribution, and then
  we extract the eigenvector matrix $\widehat{Q} \in {\mathbb R}^{p \times p}$ of
  matrix $\frac{1}{2}(\widehat{B}+\widehat{B}^\top)$. We finally set
  $B_i = \widehat{Q} \mathrm{Diag}(\widehat{c}_i) \widehat{Q}^\top$,
  where $\widehat{c}_i \in {\mathbb R}^p$ has the entries generated mutually independently
  from the interval~$[1,101]$. Each vector $a_i \in {\mathbb R}^p$
   has mutually independently generated entries from the interval~$[1,11]$.
   Note that $a_i$--the minimizer of $f_i$--is clearly known beforehand to
   node $i$, but the desired global minimizer of $f$ is not known by any node~$i$.

The logistic loss scenario corresponds to distributed learning of
 a linear classifier; see, e.g.,~\cite{BoydADMM} for details. Each node $i$ possesses $J=2$ data samples
 $\{a_{ij}, b_{ij}\}_{j=1}^{J}$. Here, $a_{ij} \in {\mathbb R}^{3}$ is a feature vector, and
$b_{ij} \in \{-1,+1\}$ is its class label.
 We want to learn a vector $x=(x_1^\top,x_0)^\top$,
$x_1 \in {\mathbb R}^{p-1}$, and
$x_0 \in {\mathbb R}$, $p \geq 2$, such that
the total logistic loss with $l_2$ regularization is minimized:
 $$
\sum_{i=1}^n \sum_{j=1}^J \mathcal{J}_{\mathrm{logis}} \left(  b_{ij} (x_1^{\top}a+x_0) \right) + \tau  \|x\|^2,
 $$
Here, $\mathcal{J}_{\mathrm{logis}}(\cdot)$ is the logistic loss $$\mathcal{J}_{\mathrm{logis}}(z) = \log (1+e^{-z}),$$
and $\tau $ is a positive regularization parameter. Note that, in this example,
we have $$f_i(x)=\sum_{j=1}^J \mathcal{J}_{\mathrm{logis}} \left(  b_{ij} (x_1^{\top}a+x_0)\right) + \frac{\tau}{n} \|x\|^2,$$
 $f(x)=\sum_{i=1}^n f_i(x)$.
 The $a_{ij}$'s are generated independently over $i$ and $j$, where each entry of $a_{ij}$ is drawn independently
from the standard normal distribution.
The ``true'' vector $x^\star=((x_1^\star)^\top, x_0^\star)^\top$ is obtained
by drawing its entries independently from standard normal distribution.
Then, the class labels are $b_{ij}=\mathrm{sign} \left( (x^\star_1)^\top a_{ij}+x^\star_0+\varepsilon_{ij}\right)$, where
$\varepsilon_{ij}$'s are drawn independently from normal distribution with zero mean and
standard deviation~$0.1$.

The network instances are generated from the random geometric graph model: nodes are placed
uniformly at random over a unit square, and the node pairs within distance~$r= \sqrt{\frac{\mathrm{ln}(n)}{n}}$ are connected with edges.
All instances of networks used in the experiments are connected.
The weight matrix $W$ is set as follows.
 For a pair of nodes $i$ and $j$ connected with an edge,
  $w_{ij} = \frac{1}{2\max\{d_i,d_j\}+1}$, where $d_i$ is the degree of the node $i$; for a pair of nodes
 not connected by an edge, we have $w_{ij}=0$; and
 $w_{ii} = 1 - \sum_{j \neq i} w_{ij}$, for all $i$.
  For the case of regular graphs, considered in~\cite{ribeiro,ribeiroNNpart1,ribeiroNNpart2},
  this weight choice coincides with that in~\cite{ribeiro,ribeiroNNpart1,ribeiroNNpart2}.

The proposed methods DQN are compared with the methods NN-$0$, NN-$1$, and NN-$2$
proposed in \cite{ribeiro}. The methods NN-$\ell$, with $\ell \geq 3$ are not numerically
tested in~\cite{ribeiro} and require a large communication cost per iteration.
 Recall that the method proposed in this paper are denoted DQN-$\ell$ with $\mathbb L_k = 0 $ as DQN-$0$;
 it has the same
  communication cost per iteration~$k$ as NN-$0$, where each
  node transmits one ($p$-dimensional) vector per iteration.
  Similarly, DQN-$1$ corresponds to
  NN-$1$, where two per-node vector communications are utilized,
  while DQN-$2$ corresponds to NN-$2$
  (3 vector communications per node).

With both the proposed methods and the methods in~\cite{ribeiro},
 the step size~$\varepsilon=1$ is used.  Step size $\varepsilon=1$
  has also been used in~\cite{ribeiro,ribeiroNNpart1,ribeiroNNpart2}.
  Note that both classes  of methods  -- NN and DQN -- guarantee global convergence with~$\varepsilon=1$
 for quadratic costs, while neither of the two groups of methods
 have guaranteed global convergence with logistic losses. For the proposed
 methods, safeguarding is not used with quadratic costs.
  With logistic costs, the safeguarding is not used
 with DQN-$0$ and $2$ but it is used with DQN-$1$, which diverges without the safeguard
 on the logistic costs. The safeguard parameter  $\rho$ defined
 as the upper bound in~(18) with $\delta=0$ is employed.
  Further, with all DQNs,
 $\theta=0$ is used.  With all the algorithms,
 each node's solution estimate is initialized by a zero vector.

The following error metric
\[
\frac{1}{n}\sum_{i=1}^n \frac{\left\| x_i^k-x^\star\right\|_2}{\|x^\star\|_2},\,\,x^\star \neq 0,
\]
is used and refered to as the relative error at iteration~$k$.

Figure~\ref{Figure_quadratic_small} (left) plots the relative error versus
the number of iterations~$k$ for a network with $n=30$ nodes,
and the quadratic costs with the variable dimension~$p=4$. First, we can see
that the proposed DQN-$\ell$ methods perform better than their corresponding counterparts NN-$\ell$,
$\ell=0,1,2$. Also, note that the performance of DQN-1 and DQN-2 in terms of
iterations match in this example.
 Figure~\ref{Figure_quadratic_small} (right)
 plots the relative error versus total number of communications.
 We can see that, for this example, DQN-$0$ is the most
  efficient among all methods in terms of the communication cost.
  Further, interestingly, the performance of NN-$0$, NN-$1$, and NN-$2$ is
  practically the same in terms of communication cost on this example. The clustering
  of the performance of NN-$0$, NN-$1$, and NN-$2$ (although not so pronounced as in our examples)
  emerges also in Simulations in~\cite{ribeiro,ribeiroNNpart1,ribeiroNNpart2}.
  Also, the performance of DQN-0 and NN-1
  practically matches.
   In summary, method DQN-$0$ shows the best performance
  in terms of communication cost on this example, while
  DQN-$1$ and $2$ are the best in terms of the number of iterations~$k$.

The improvements
of DQN over NN are mainly due to
the different splitting parameter $\theta=0$.
Actually,
  our numerical experience suggests
  that NN-$\ell$ may perform
  better than DQN-$\ell$ with $\theta=1$
  for $\ell=1,2.$
 We provide an intuitive explanation for the advantages of choice $\theta=0$,
focusing on the comparison
between NN-0 and DQN-0.
 Namely,
   the adopted descent directions
   with both of these methods
   correspond to the quality of the zeroth
   order Taylor expansion of
   the following matrix:
\begin{equation}
\label{eqn-taylor-approx-0}
\left( \mathbb{I} - (\mathbb{A}_k(\theta))^{-1} \mathbb{G}\right)^{-1} \approx \mathbb{I}.
\end{equation}
%
 In~(\ref{eqn-taylor-approx-0}),
 with NN-0, we have
 $\mathbb{A}_k(\theta) = \mathbb{A}_k(\theta=1)$,
 while with DQN-0, we have that:
 $\mathbb{A}_k(\theta) = \mathbb{A}_k(\theta=0)$; note that
 these two matrices are different.
 Now, the error (remainder)
 of the Taylor approximation is roughly of size $\|(\mathbb{A}_k(\theta))^{-1} \mathbb{G}\|$.
  In view of the upper bound in~(\ref{eqn-spec-rad-GAinv}),
   we have with NN-0 that the remainder is of size:
   \begin{equation}
   \label{eqn-explanation}
   \|(\mathbb{A}_k(1))^{-1} \mathbb{G}\| \approx 1-\frac{\alpha \,\mu}{2(1-w_{min})},
   \end{equation}
  for small $\alpha \,\mu$.
  On the other hand, we have that the DQN-0's remainder is:
  \[
   \|(\mathbb{A}_k(0))^{-1} \mathbb{G}\| \approx 1-\frac{\alpha \,\mu}{1-w_{min}}.
   \]
  Therefore,
  the remainder is (observed through this rough, but indicative, estimate)
   larger with NN-0,
   and that is why DQN-0 performs better.
  We can similarly compare
  NN-1 (which corresponds
  to the first order Taylor approximation
  of the matrix in~(49)) with DQN-0.
   The remainder with NN-1
   is roughly
   $\|(\mathbb{A}_k(1))^{-1} \mathbb{G}\|^2 \approx 1-\frac{\alpha \,\mu}{1-w_{min}}$,
   which equals to the remainder
   estimate of DQN-0. This
   explains why the two methods perform
   very similarly.
    Finally, note that
the upper bound on~$\|\mathbb{G} {\mathbb A}_k^{-1}\|$
in~(\ref{eqn-spec-rad-GAinv})
is an increasing function of $\theta \geq 0$
 (the lower the $\theta$, the better the bound),
 which justifies the choice $\theta=0$
  adopted here for DQN.


Figure~\ref{Figure_quadratic_large} (left and right) repeats the plots
for the network with~$n=400$ nodes, quadratic costs, and the variable dimension
 $p=3$. One can see that again the proposed methods outperform their
 respective NN-$\ell$ counterparts. In terms of communication cost,
 DQN-$0$ and DQN-$1$ perform practically the same and are the most efficient among all methods.

\begin{figure}[thpb]
      \centering
      \includegraphics[height=2.4 in,width=2.3 in]{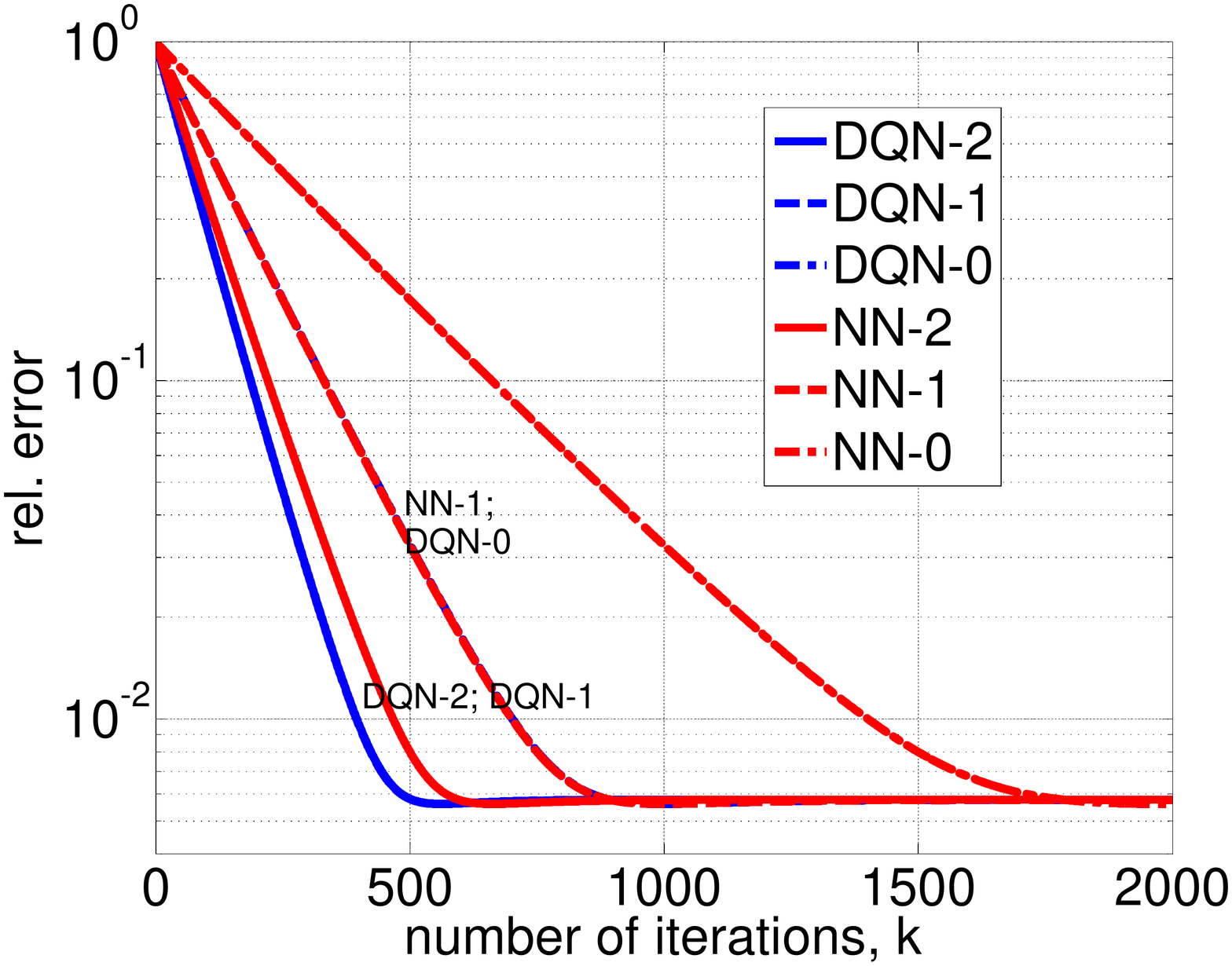}
      \includegraphics[height=2.4 in,width=2.3 in]{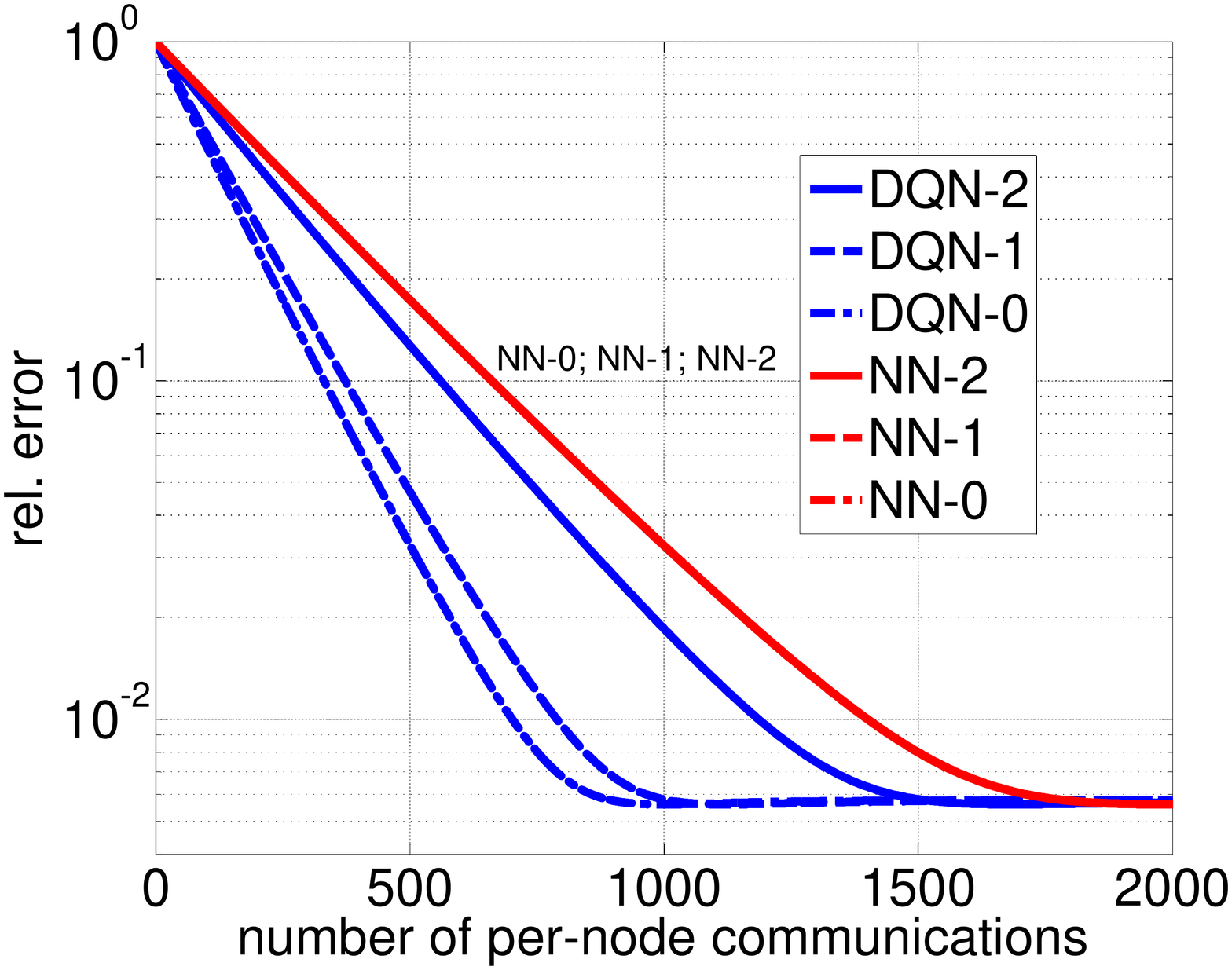}
      \caption{Relative error versus number of iterations~$k$ (left) and versus number of communications (right) for
      quadratic costs and $n=30$-node network. }
      \label{Figure_quadratic_small}
\end{figure}

\begin{figure}[thpb]
      \centering
      \includegraphics[height=2.4 in,width=2.3 in]{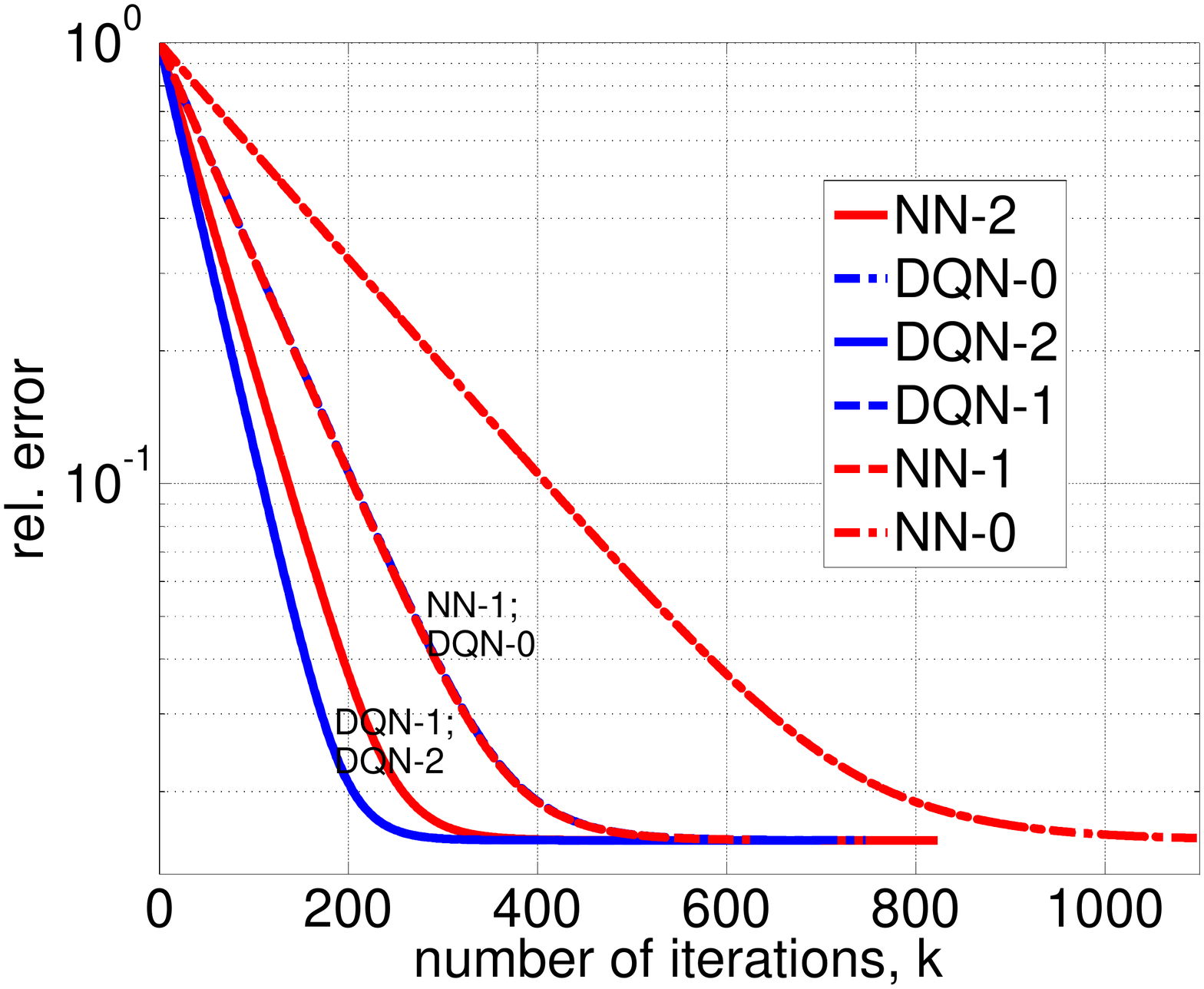}
      \includegraphics[height=2.4 in,width=2.3 in]{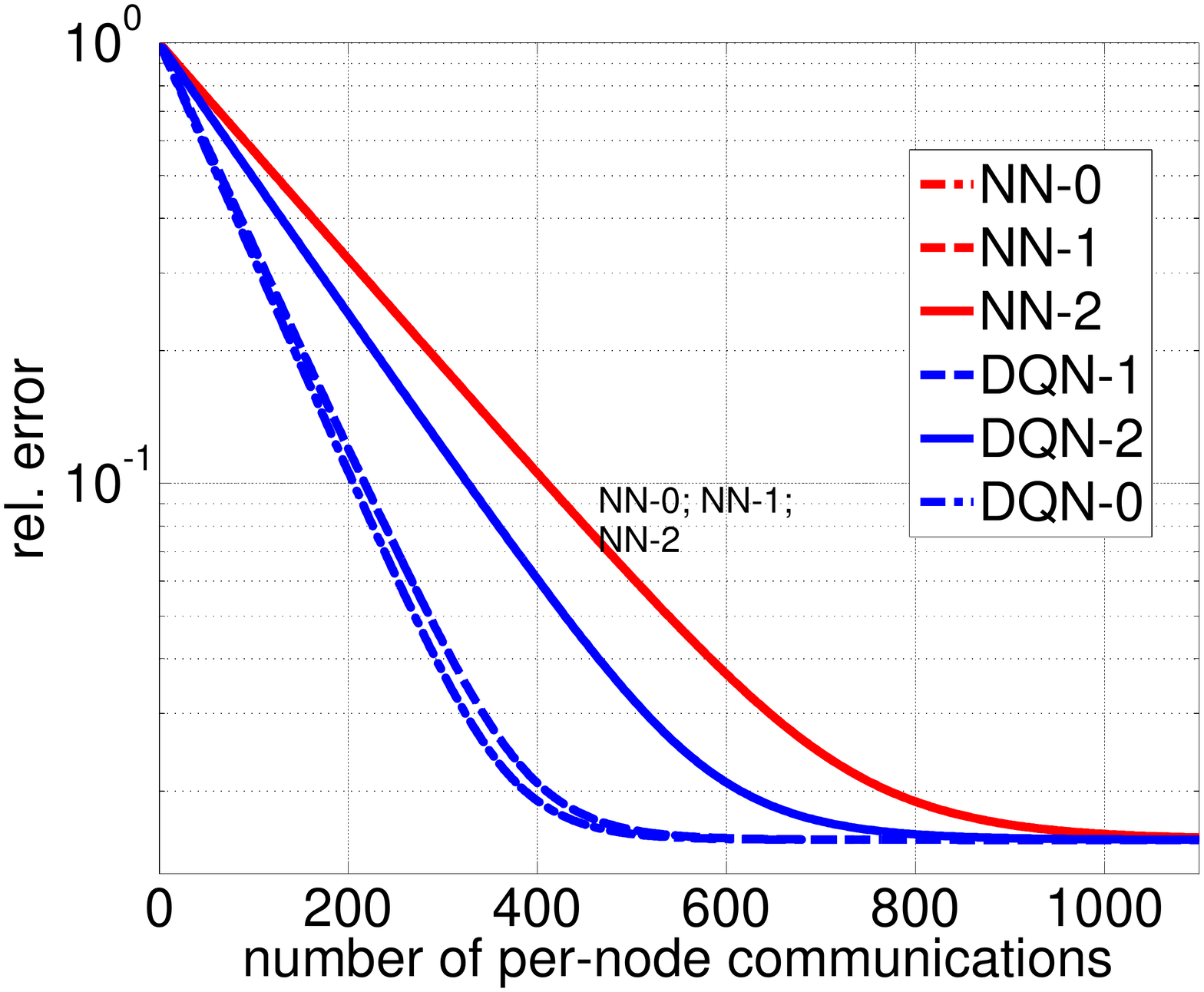}
      \caption{Relative error versus number of iterations~$k$ (left) and versus number of communications (right)
      for
      quadratic costs and $n=400$-node network. }
      \label{Figure_quadratic_large}
\end{figure}

Figure~\ref{Figure_logistic_small} plots the relative error versus number of iterations (left)
 and number of per-node communications (right) for the logistic losses with
 variable dimension $p=4$ and the network with $n=30$ nodes.
 One can see that again the proposed methods perform better than
 the NN-$\ell$ counterparts. In terms of the communication cost,
 DQN-$0$ is the most efficient among all methods,
 while DQN-$2$ is fastest in terms of the number of iterations. Finally,
 Figure~\ref{Figure_logistic_large} repeats the plots for
 variable dimension $p=4$ and the network with $n=200$ nodes, and it
 shows similar conclusions: among all DQN and NN methods, DQN-$0$ is the most efficient in terms of communications,
 while DQN-$2$ is fastest in terms of the number of iterations.

\begin{figure}[thpb]
      \centering
      \includegraphics[height=2.4 in,width=2.3 in]{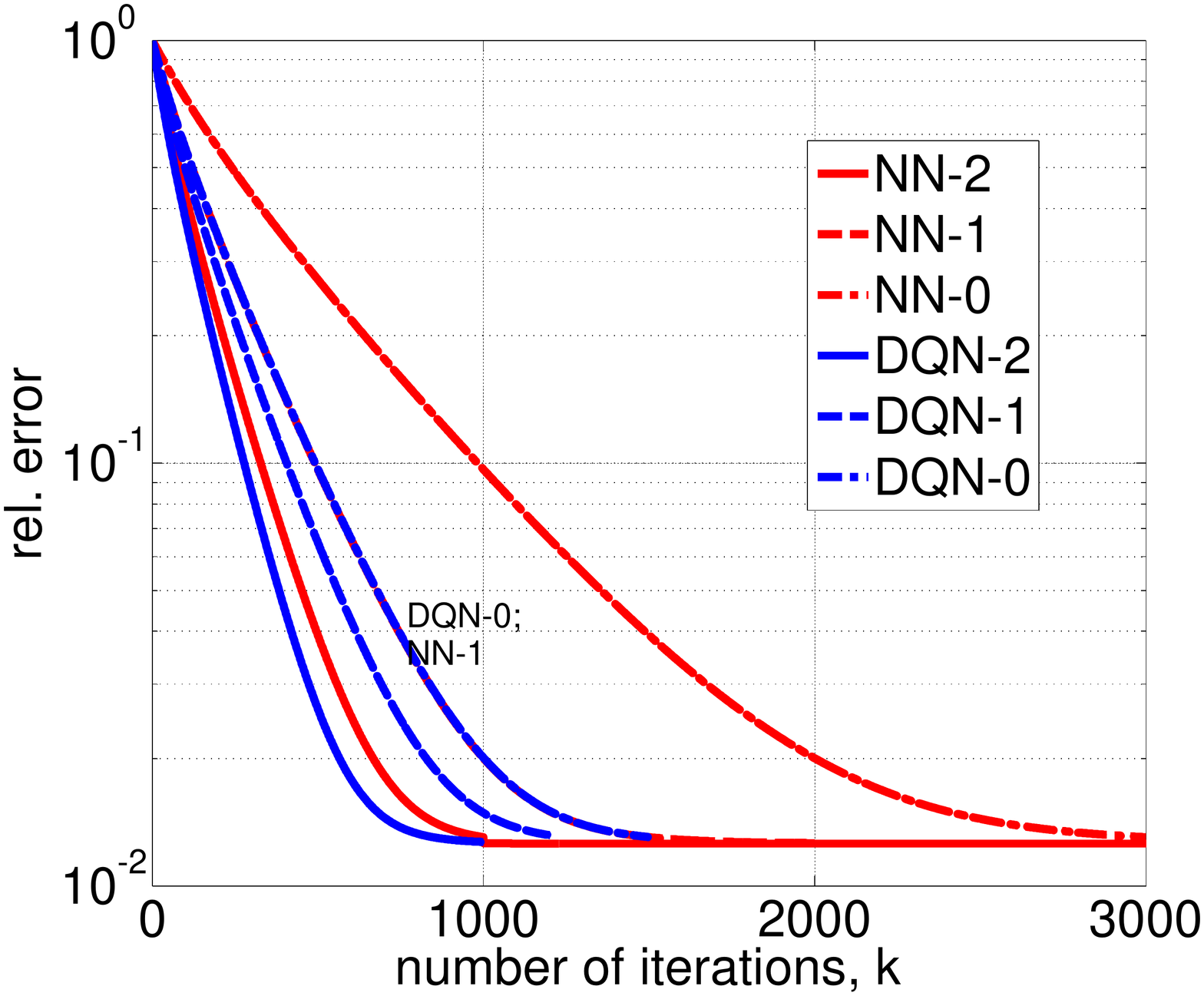}
      \includegraphics[height=2.4 in,width=2.3 in]{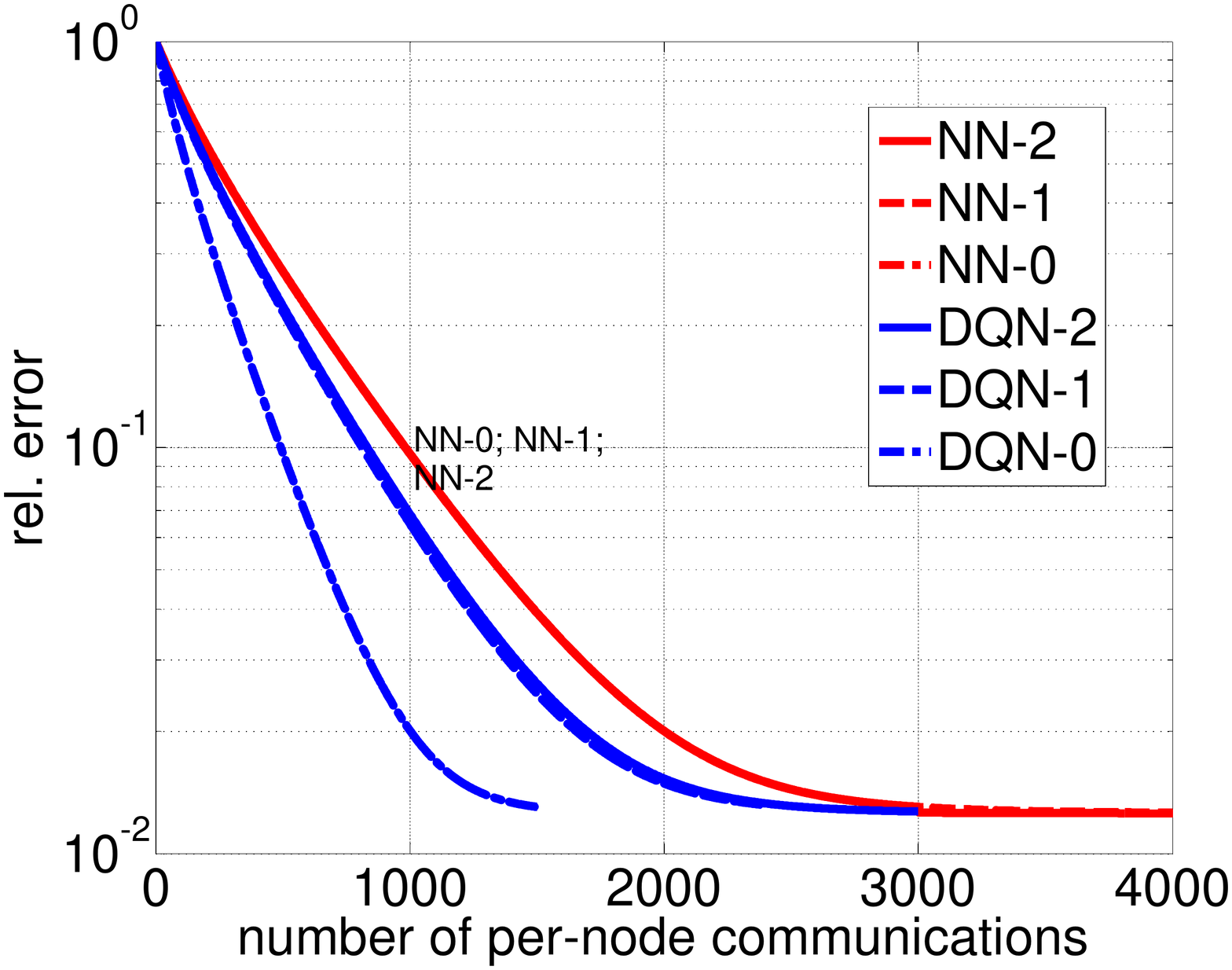}
      \caption{Relative error versus number of iterations~$k$ (left) and versus number of communications (right)
      for
      logistic costs and $n=30$-node network. }
      \label{Figure_logistic_small}
\end{figure}

\begin{figure}[thpb]
      \centering
      \includegraphics[height=2.4 in,width=2.3 in]{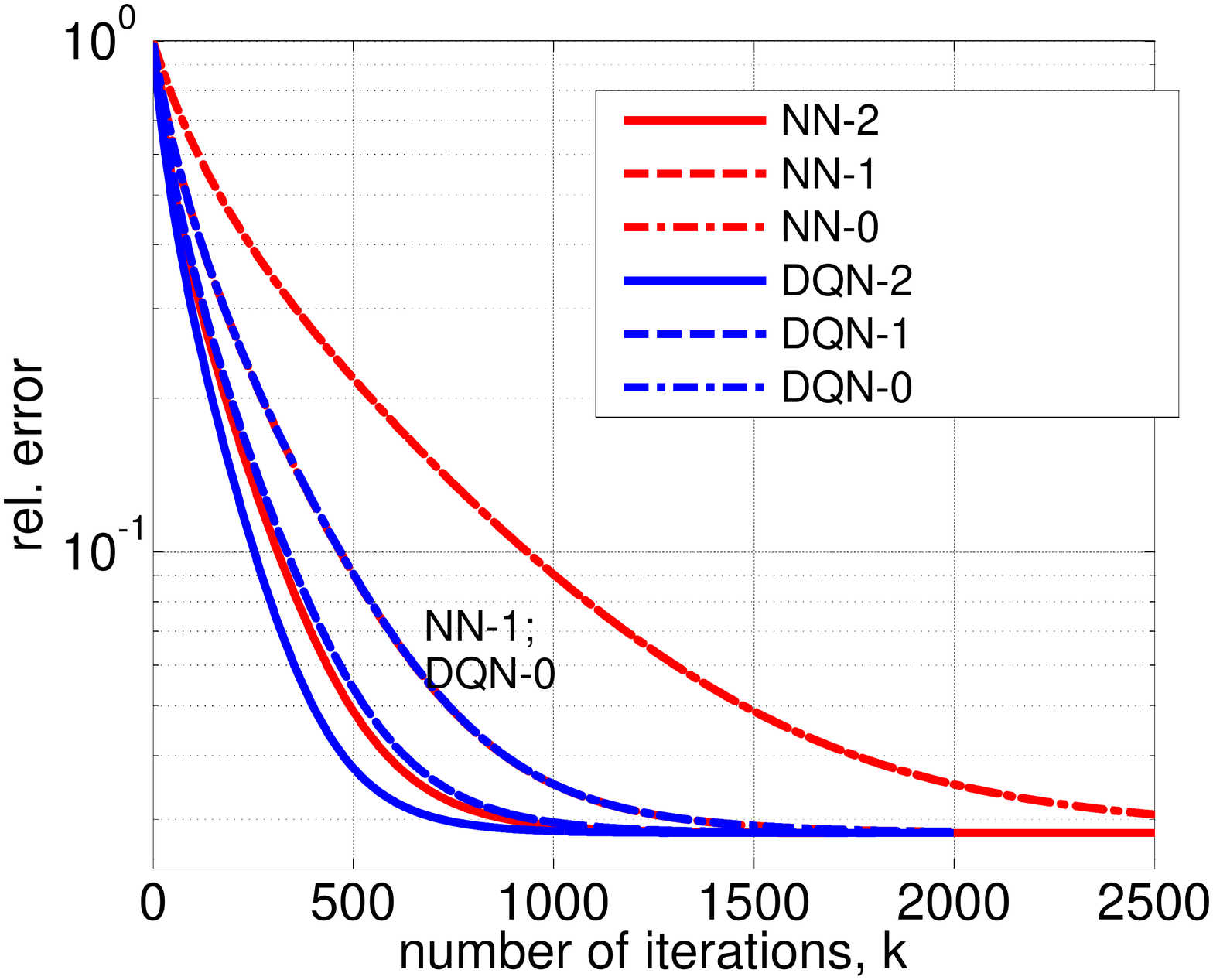}
      \includegraphics[height=2.4 in,width=2.3 in]{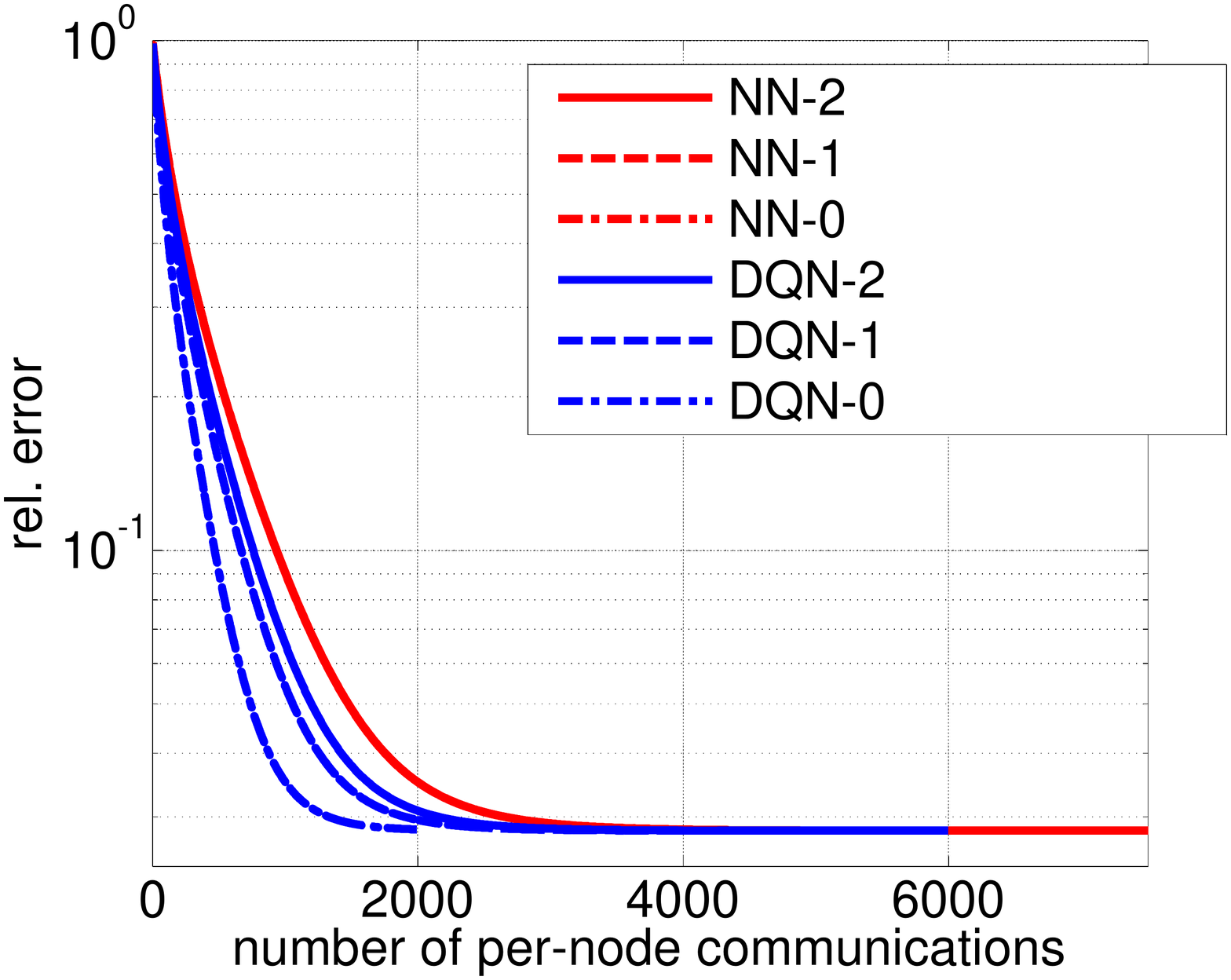}
      \caption{Relative error versus number of iterations~$k$ (left) and versus number of communications (right)
      for
      logistic costs and $n=200$-node network. }
      \label{Figure_logistic_large}
\end{figure}

\section{Extensions}

As noted before,
DQN methods do not converge to the
exact solution of~(1) but to
a solution neighborhood controlled
by the step size~$\alpha$.
As such, for high solution accuracies required, they
may not be competitive with
distributed second order methods
which converge to the exact solution~\cite{ESOM,NewtonRaphsonConsensus,DQM}.

However, we
can exploit the results of~\cite{ESOM}
and ``embed'' the DQN algorithms
in the framework of proximal multiplier methods~(PMMs),
just like~\cite{ESOM} embeds
the NN methods into the PMM framework.
 We refer to the resulting
 algorithms
 as PMM-DQN-$\ell$,
 $\ell=0,1,2.$
 (Here, PMM-DQN-$\ell$
 parallels DQN-$\ell$
 in terms of complexity of approximating
 Hessians, and in terms
 of the communication cost per iteration.)
 It is worth noting that the
 contribution to embed
 distributed second order methods
 into the PMM framework is due~\cite{ESOM}.
  Here we extend~\cite{ESOM}
  to demonstrate (by simulation)
 that the DQN-type
 Hessian approximations
 within the PMM framework
  yield efficient
  distributed second order methods.

We now briefly describe a general PMM-DQN method;
for the methodology to devise
distributed second order PMM methods,
we refer to~\cite{ESOM}. See also the Appendix for further details.
  We denote
 by $\widehat{x}^{\,k} =
 \left( \widehat{x}^{\,k}_1,...,\widehat{x}^{\,k}_n\right)
  \in {\mathbb R}^{n p}$
   the current iterate,
   where $\widehat{x}^{\,k}_i \in {\mathbb R}^p$
    is node $i$'s estimate of
    the solution to~(1) at iteration~$k$.
    Besides $\widehat{x}^{\,k}$ (the primal variable),
    the PMM-DQN method
    also maintains a dual variable
    $\widehat{q}^{\,k}=\left(
    \widehat{q}^{\,k}_1,...,\widehat{q}^{\,k}_n\right) \in {\mathbb R}^{n\,p}$,
    where $\widehat{q}^{\,k}_i \in {\mathbb R}^p$
     is node~$i$'s
      dual variable at iteration~$k$.
      Quantities
      $\widehat{\mathbb H}_k$,
      $\widehat{\mathbb A}_k$,
      $\widehat{\mathbb G}$ and
      $\widehat{g}_k$ defined
      below play a role in the PMM-DQN
      method and
      are, respectively,  the counterparts
      of $\nabla^2 \Phi(x^{\,k})$,
      ${\mathbb A}_k$,
      ${\mathbb G}$ and
      $\nabla \Phi(x^{\,k})$ with DQN:
      \begin{eqnarray}
      \label{eqn-new-Hessian}
      \widehat{\mathbb H}_k &=& \nabla^2 F(\widehat{x}^{\,k})
       + \beta\,\left( \II - \mathbb Z \right) + \epsilon_{\mathrm{pmm}}\,\II  = \widehat{\mathbb A}_k - \widehat{\mathbb G}\\
       \label{eqn-new-splitting-A-k}
      \widehat{\mathbb A}_k &=&
      \nabla^2 F(\widehat{x}^{\,k})
       + \beta\,\left( \II - {\mathbb Z}_d \right) + \epsilon_{\mathrm{pmm}}\,\II + \beta\,\theta\left( \II - \mathbb{Z}_d \right)\\
        \label{eqn-new-splitting-G}
        \widehat{\mathbb G}
        &=&
      \beta\,{\mathbb Z}_u + \beta\,\theta\,\left(\II - \mathbb{Z}_d\right)\\
      \widehat{g}_k &=& \nabla F(\widehat{x}^{\,k}) + \beta\,(\II-\mathbb{Z})\widehat{x}^{\,k} + \widehat{q}^{\,k}.
      \end{eqnarray}
Here, $\theta\geq 0$
 is the splitting parameter as with DQN, $\beta>0$ is the dual step size
and $\epsilon_{\mathrm{pmm}}>0$
relates to the proximal term of
the corresponding augmented Lagrangian; see~\cite{ESOM}
for details.
We now present the
general PMM-DQN algorithm.
Note from Step~2 the analogous form
of the Hessian inverse approximation as with DQN;
the approximation is again
parameterized with a $(n p) \times (n p)$
 diagonal matrix~$\widehat{\mathbb L}_k.$

\vspace{5mm}

 \noindent{\bf Algorithm 4: PMM-DQN in vector format} \\
 Given $ x^0 =0$, $ \beta, \epsilon_{\mathrm{pmm}}, \rho > 0,\; \theta \geq 0 $.
 Set $ k = 0. $
 \begin{itemize}
 \item[Step 1.] Chose a diagonal matrix $\widehat{\LL}_{k} \in \mathbb{R}^{np \times np} $ such that $$\|\widehat{\LL}_{k}\| \leq \rho.$$
 \item[Step 2.]
Set
$$  \widehat{s}^{\,k} =
- (\II - \widehat{\LL}_k \widehat{\GG}) \widehat{{\mathbb A}}_k^{-1} \,\widehat{g}_k. $$
 \item[Step 3.] Set
 $$ \widehat{x}^{\,k+1} = \widehat{x}^{\,k} + \widehat{s}^{\,k}.$$
 \item[Step 4.] Set
 $$\widehat{q}^{\,k+1}  = \widehat{q}^{\,k} + \left( \II - \mathbb{Z} \right)\widehat{x}^{\,k+1};\; k = k+1. $$
 \end{itemize}
The same algorithm is presented below
from the distributed implementation perspective.
 (Note that here we adopt the notation
similar to DQN, i.e.,
the $(i,j)$-th
$p \times p$
 block of $\widehat{\mathbb G}$
  is denoted by $\widehat{G}_{ij}$;
  the $i$-th $p \times p$
   diagonal block of
   $\widehat{\mathbb L}_k$ is denoted by
   $\widehat{\Lambda}^{\,k}_i$;
   and $i$-th $p \times p$
    diagonal block of $\widehat{\mathbb A}_k$
     is denoted by $\widehat{A}^{\,k}_i$.)

\vspace{5mm}
\noindent{\bf Algorithm 5: PMM-DQN -- Distributed implementation}\\
\noindent{At each node $i$, require $\beta,\rho,\epsilon_{\mathrm{pmm}}>0 $, $\theta \geq 0$.}
\begin{itemize}
\item[1] Initialization: Each node $i$ sets $k=0$ and $\widehat{x}_i^0 = \widehat{q}_i^0 = 0$.
    \item[2] Each node $i$ calculates
    \[
    \widehat{d}_i^{\,k} =
    \left( \widehat{A}_i^{\,k}\right)^{-1} \left[ \, \nabla f_i(\widehat{x}_i^{\,k})
    + \beta\,\sum_{j \in O_i} w_{ij} \left( \widehat{x}_i^{\,k} - \widehat{x}_j^{\,k}\right) + \widehat{q}_i^{\,k} \,\right].
    \]
    \item[3] Each node~$i$ transmits $\widehat{d}_i^{\,k}$ to all its neighbors
    $j \in O_i$ and receives $\widehat{d}_j^{\,k}$ from all $j \in O_i$.
    \item[4] Each node $i$ chooses a diagonal $p \times p$ matrix $\widehat{\Lambda}_i^{\,k}$, such that
    $ \|\widehat{\Lambda}_i^{\,k}\| \leq \rho. $
    \item[6] Each node $i$ calculates:
    \[
    \widehat{s}_i^{\,k}  = - \widehat{d}_i^{\,k} +
    \widehat{\Lambda}_i^{\,k} \sum_{j \in \bar{O}_i} \widehat{G}_{ij} \,\widehat{d}_j^{\,k}.
    \]
    \item[6] Each node $i$ updates its solution estimate as:
    \[
    \widehat{x}_i^{\,k+1} = \widehat{x}_i^{\,k} + \widehat{s}_i^{\,k}.
    \]
    \item[7] Each node $i$ transmits $\widehat{x}_i^{\,k+1}$ to all its neighbors
    $j \in O_i$ and receives $\widehat{x}_j^{\,k+1}$ from all $j \in O_i$.
    \item[8] Each node~$i$ updates the dual variable
    $\widehat{q}_i^{k}$ as follows:
    \[
    \widehat{q}_i^{\,k+1} = \widehat{q}_i^{\,k} + \sum_{j \in O_i} w_{ij} \left( \widehat{x}_i^{\,k+1} - \widehat{x}_j^{\,k+1} \right).
    \]
    \item[9] Set $k=k+1$ and go to Step~2.
\end{itemize}

%
Note that, at Step 2, each node $i$ needs
the neighbors' estimates $\widehat{x}_j^{k}$, $j \in O_i$. For $k \geq 1$, the availability of such information is ensured through Step~7 of the previous iteration~$k-1$; at $k=0$, Step 2 is also realizable as it is assumed that
$\widehat{x}_i^0=0$, for all $i=1,...,n$.
As with the DQN methods,
different variants PMM-DQN-$\ell$
 mutually differ in the choice of matrix~$\widehat{\LL}_k$.
 With PMM-DQN-$0$, we set $\widehat{\LL}_k=0$;
 with PMM-DQN-$2$, $\widehat{\LL}_k$ is set as described below;
 with PMM-DQN-$1$,
 we set $\widehat{\LL}_k=\widehat{\LL} = \mathrm{const}$
  to $\widehat{\LL}_0$, i.e.,
  to the value of $\widehat{\LL}_k$
  from the first iteration of PMM-DQN-$2$.

We now detail how $\widehat{\mathbb L}_k$
 is chosen with PMM-DQN-$2$.
 The methodology is completely analogous
 to DQN-$2$:
 the idea is
 to approximate the exact Newton direction~$\widehat{s}_N^k$
  which obeys the following Newton-type equation:
 \begin{equation}
 \label{eqn-secant-new}
 \widehat{\mathbb H}_k\,\widehat{s}^{\,k} + \widehat{g}_k = 0,
 \end{equation}
  through Taylor expansions.
  Using~\eqref{eqn-secant-new}
  and completely analogous steps
  as with DQN-$2$, it follows
  that~$\widehat{\mathbb L}_k$
   is obtained through solving
   the following system of linear equations with respect
   to~$\widehat{\mathbb L}_k$:
\begin{eqnarray}
&\,&\widehat{\mathbb L}_k\, \widehat{u}^{\,k} = - \left( \frac{1}{\beta+\epsilon_{\mathrm{pmm}}}
\II + \frac{\beta}{(\beta+\epsilon_{\mathrm{pmm}})^2} \mathbb{Z} - \frac{1}{(\beta+\epsilon_{\mathrm{pmm}})^2} \nabla^2 F(\widehat{x}^k)\right)  \widehat{u}^{\,k} \nonumber \\
&\,&
\label{eqn-hat-u-k-def}
\mathrm{where} \,\,\,\,\, \widehat{u}^{\,k}
 = \widehat{\mathbb G}\,{\widehat{\mathbb A}}_k^{\,-1}\,\widehat{g}_k.
\end{eqnarray}
The overall algorithm for
setting
$\widehat{\mathbb L}_k$
 with PMM-DQN-$2$ (step~4 of Algorithm~5) is presented below.

  \vspace{5mm}

\noindent{\bf Algorithm~6: Computing $\widehat{\mathbb L}_k$ with PMM-DQN-$2$}
\begin{itemize}
    \item[4.1] Each node~$i$ calculates
    \[
    \widehat{u}_i^{\,k} = \sum_{j \in \bar{O}_i}\widehat{G}_{ij}\,\widehat{d}_j^{\,k}.
    \]
    \item[4.2] Each node~$i$ transmits $\widehat{u}_i^{\,k}$ to all its neighbors $j \in O_i$
     and receives $\widehat{u}_j^{\,k}$ from all $j \in O_i$.
    \item[4.3] Each node $i$ calculates $\widehat{\Lambda}_i^{\,k}$ -- the solution to the following
    system of equations (where the only unknown is the $p \times p$
     diagonal matrix $\widehat{\Lambda}_{i}$):
    \begin{eqnarray*}
    &\,& \widehat{\Lambda}_i \,\widehat{u}_i^{\,k} =
    - \left[\,
    (\frac{1}{\beta+\epsilon_{\mathrm{pmm}}} +
     \frac{\beta\,w_{ii}}{(\beta+\epsilon_{\mathrm{pmm}})^2}) I-\frac{1}{(\beta+\epsilon_{\mathrm{pmm}})^2}\nabla^2 f_i(\widehat{x}_i^{\,k})\,\right] \\
    &\,& \phantom{\Lambda_i \,\widehat{u}_i^{\,k}=} \times \widehat{u}_i^{\,k} -
    \frac{\beta}{(\beta+\epsilon_{\mathrm{pmm}})^2} \sum_{j \in O_i}w_{ij}\,\widehat{u}_j^{\,k}.
    \end{eqnarray*}
    \item[4.4] Each node~$i$ projects each diagonal entry of $\widehat{\Lambda}_i^k$ onto the interval~$[-\rho,\,\rho]$.

\end{itemize}

\textbf{Simulations}. We now compare by simulation the PMM-DQN-$\ell$ methods
with the ESOM-$\ell$
algorithms proposed in~\cite{ESOM},
 the DQM algorithm in~\cite{DQM},
 and different variants of the Newton Raphson Consensus~(NRC)
 algorithm proposed in~\cite{NewtonRaphsonConsensus}.

The simulation setup is as follows.
The network is an instance of the random geometric graph
with $n=30$ nodes and $166$ links.
The optimization variable dimension is~$p=4$,
and the local nodes costs are strongly convex quadratic,
generated at random in the same way
as with the quadratic costs examples in Section~5.
 With all methods which involve weighted averaging
(ESOM-$\ell$, PMM-DQN-$\ell$, and NRC), we use the Metropolis
weights.
We consider the ESOM-$\ell$
 and PMM-DQN-$\ell$ methods
 for $\ell=0,1,2.$
  With the methods ESOM-$\ell$
  and PMM-DQN-$\ell$, we set
  the proximal constant $\epsilon_{\mathrm{pmm}}=10$
  (see~\cite{ESOM} for details).
  Further,
  we tune the dual step size $\beta$ separately for each of these
    methods to the best by
    considering the following candidate values:
     $\beta \in \{10^{-4},10^{-3.5},10^{-3},...,10^{3.5},10^4\}$,
     i.e.,
     a grid of points equidistant on the $\mathrm{log}_{10}$
      scale with the half-decade spacing.
       The algorithm
       DQM has the tuning step size parameter
        $c>0$ (see~\cite{DQM} for details) which
         we also tune to the best
         using the same grid of candidate values.
          Regrading
          the methods proposed in~\cite{NewtonRaphsonConsensus},
          we consider both
          the standard (NRC)
           and the accelerated (FNRC)
            algorithm variant.
            These methods
            have a communication
            cost per node, per iteration which is
            quadratic in~$p$, due to exchanging local Hessian
            estimates. (Specifically, as it is sufficient to transmit
            the upper-triangular part of a local Hessian (due to the matrix symmetry),
            each node per iteration transmits $p \times (p+1)/2$ scalars for the Hessian exchange.)
            This is different from
            the ESOM, DQM, and PMM-DQN methods
            which have a cost linear in~$p$.
            We also consider the Jacobi
            variants in~\cite{NewtonRaphsonConsensus} (both the standard -- JC
             and the accelerated  -- FJC variant)
             which approximate local Hessians
            through diagonal matrices
            and hence their communication cost per node,
            per iteration reduces to a cost linear in~$p$.
           With NRC, FNRC, JC, and FJC,
           we set their
         step size~$\epsilon$
          to unity (see~\cite{NewtonRaphsonConsensus} for details),
          as this (maximal possible)
           step size value yielded
           fastest convergence.
            We observed that JC and FJC
            converged with a non-zero limiting error,
           while decreasing the value of $\epsilon$
             did not improve the
            limiting error of the method while slowing down the methods.
           Hence,
           with all the methods
           considered,
            we tune their step sizes
            to the best (up to
            the finite candidate grid points resolution).
            With FNRC and FJC, we set the acceleration parameter $\phi$ in the same way as in \cite{NewtonRaphsonConsensus} (see page 10 of \cite{NewtonRaphsonConsensus}.) With all PMM-DQN methods,
            we do not use safeguarding ($\rho=+\infty$),
            and we set $\theta=0$.
            With all the methods considered, the primal variables -- solution estimates (and dual variables, if exist) are initialized with zero vectors.
             The error metric is the same as with the quadratic example in Section~5.

Figure~5 compares the
      PMM-DQN-$\ell$ algorithms
      and the ESOM-$\ell$ algorithms in~\cite{ESOM} in terms
      of the number of iterations (left)
       and the number of per-node communications (right).
       We can see that, in terms
       of iterations,
       for each fixed $\ell$,
       the corresponding PMM-DQN
       method performs better than the
       corresponding ESOM method.
       The exception is the case
       $\ell=0$ where the two methods are comparable.
       The same conclusions
       hold for
       the number of
       communications also.
       Further, in terms of the
       number of iterations, PMM-DQN-1
        and PMM-DQN-2 are the best among all methods;
        in terms of communications,
        PMM-DQN-1 is the best method
         among all PMM-DQN and ESOM
         method considered.

         In Figure~6,
         we compare the best among
         the PMM-DQN-$\ell$
          methods, the best
          among the ESOM-$\ell$
           methods (in
           terms of iterations, the best are PMM-DQN-1 and ESOM-2,
           while in terms of communications, the best are PMM-DQN-1 and ESOM-0),
           DQM, and the NRC methods group (NRC, FNRC, JC, and FJC).
           We can see that the PMM-DQN-1 method
           converges faster than all other methods,
           both in terms of iterations and communications.

\begin{figure}[thpb]
      \centering
      \includegraphics[height=2.4 in,width=2.3 in]{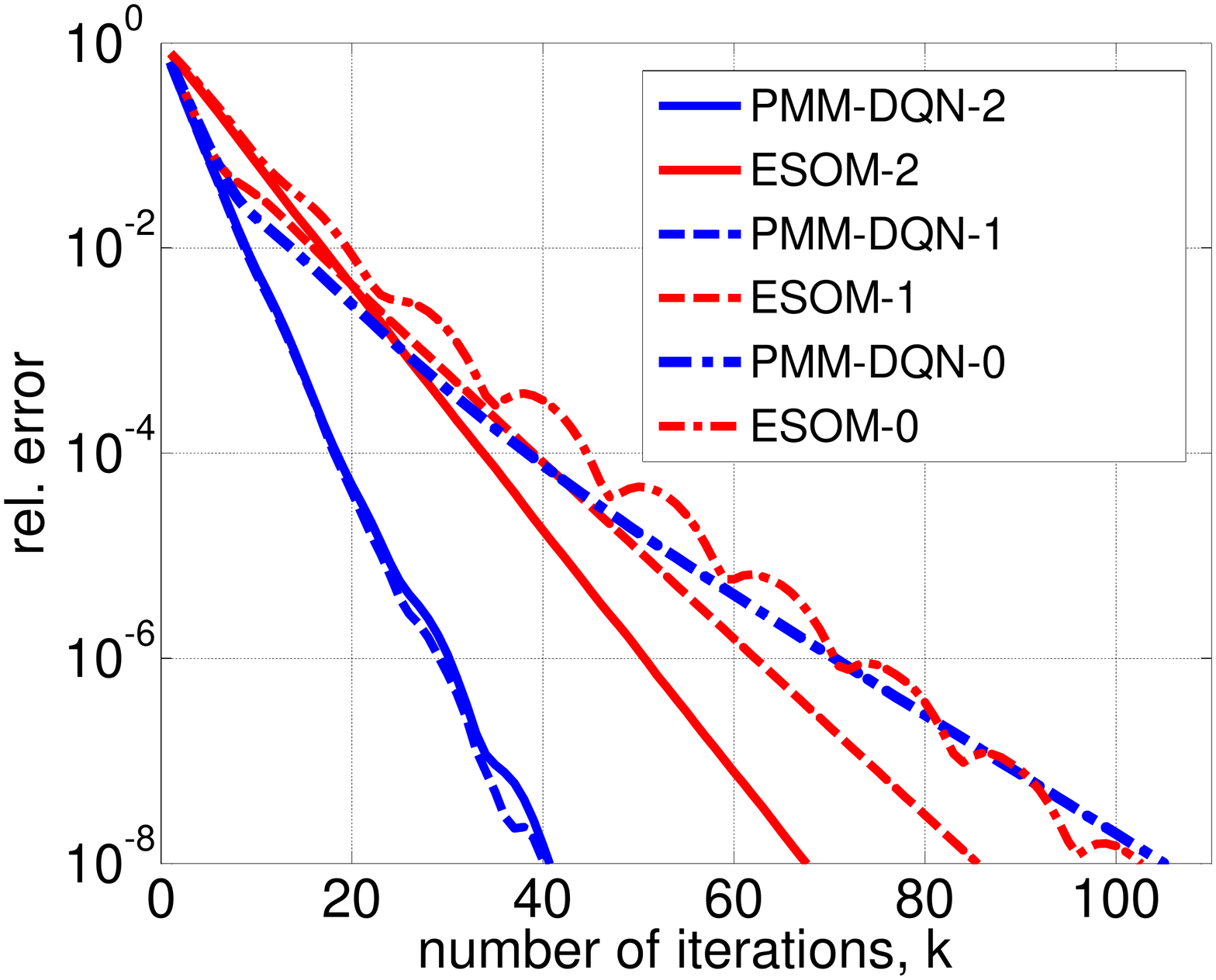}
      \includegraphics[height=2.4 in,width=2.3 in]{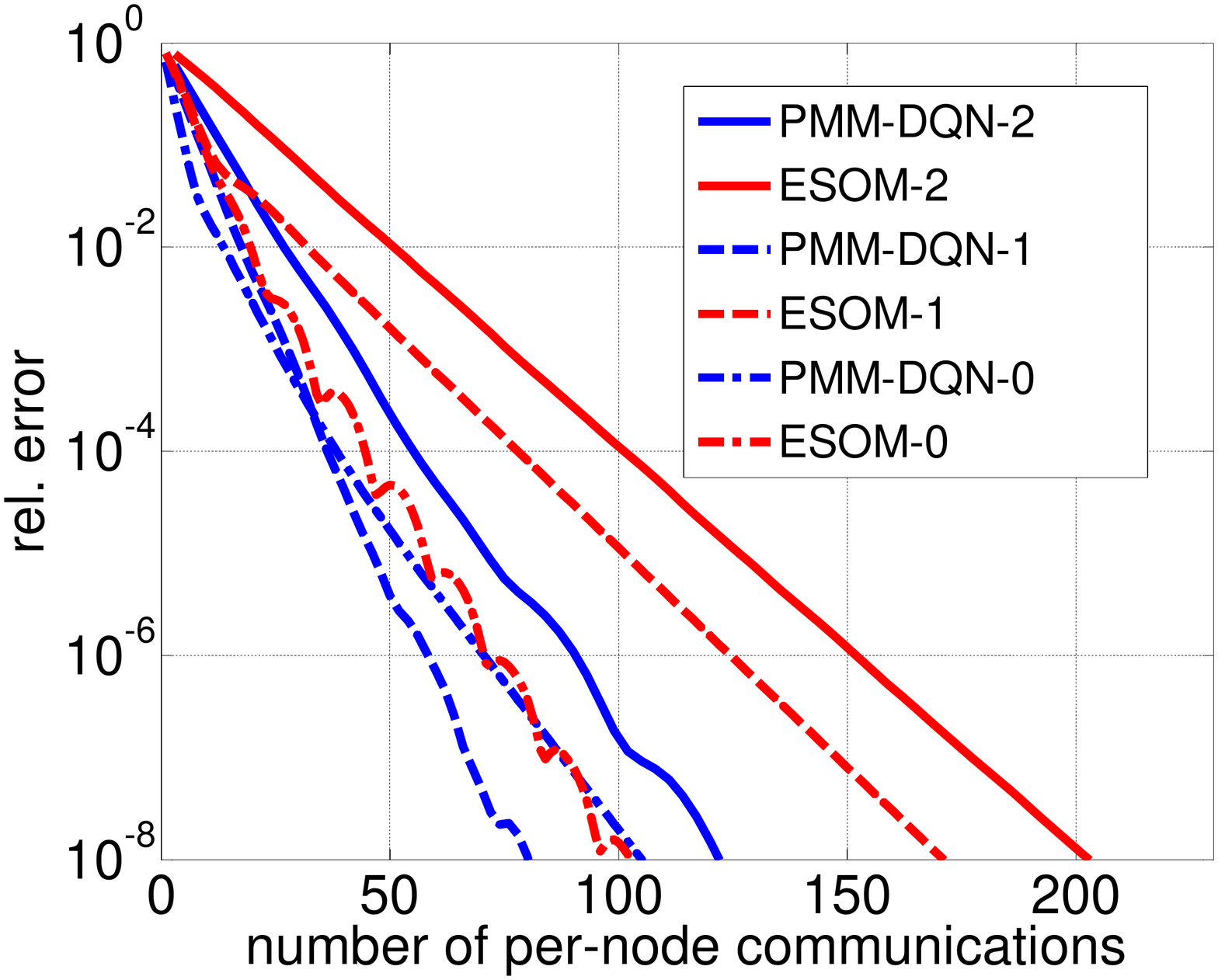}
      \caption{Comparison between the
      PMM-DQN-$\ell$ algorithms
      and the ESOM-$\ell$ algorithms in~\cite{ESOM}. The figures plot relative error versus number of iterations~$k$ (left) and versus number of communications (right)
      for quadratic costs and $n=30$-node network.}
      \label{Figure_ESOM_1}
\end{figure}

\begin{figure}[thpb]
      \centering
      \includegraphics[height=2.4 in,width=2.3 in]{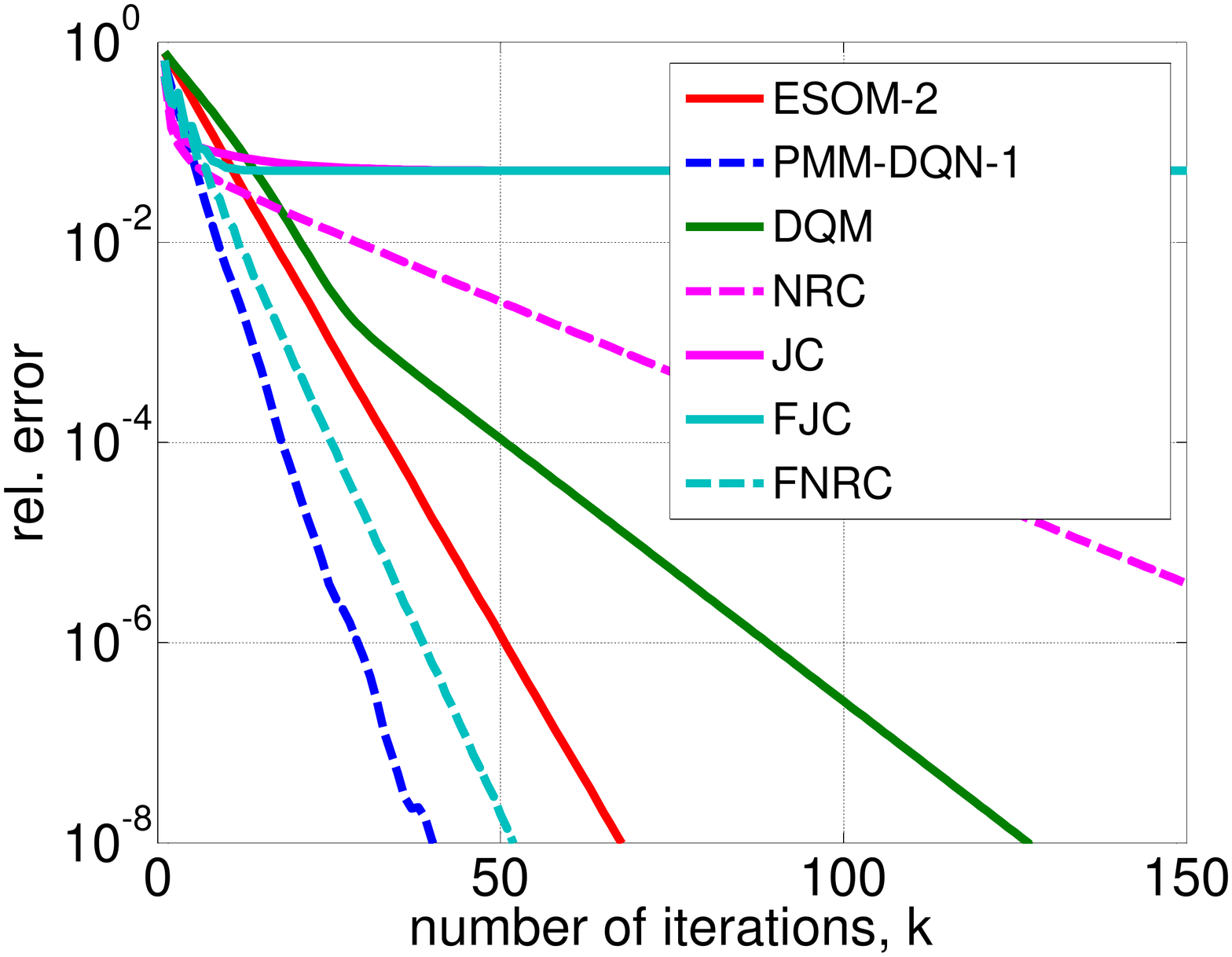}
      \includegraphics[height=2.4 in,width=2.3 in]{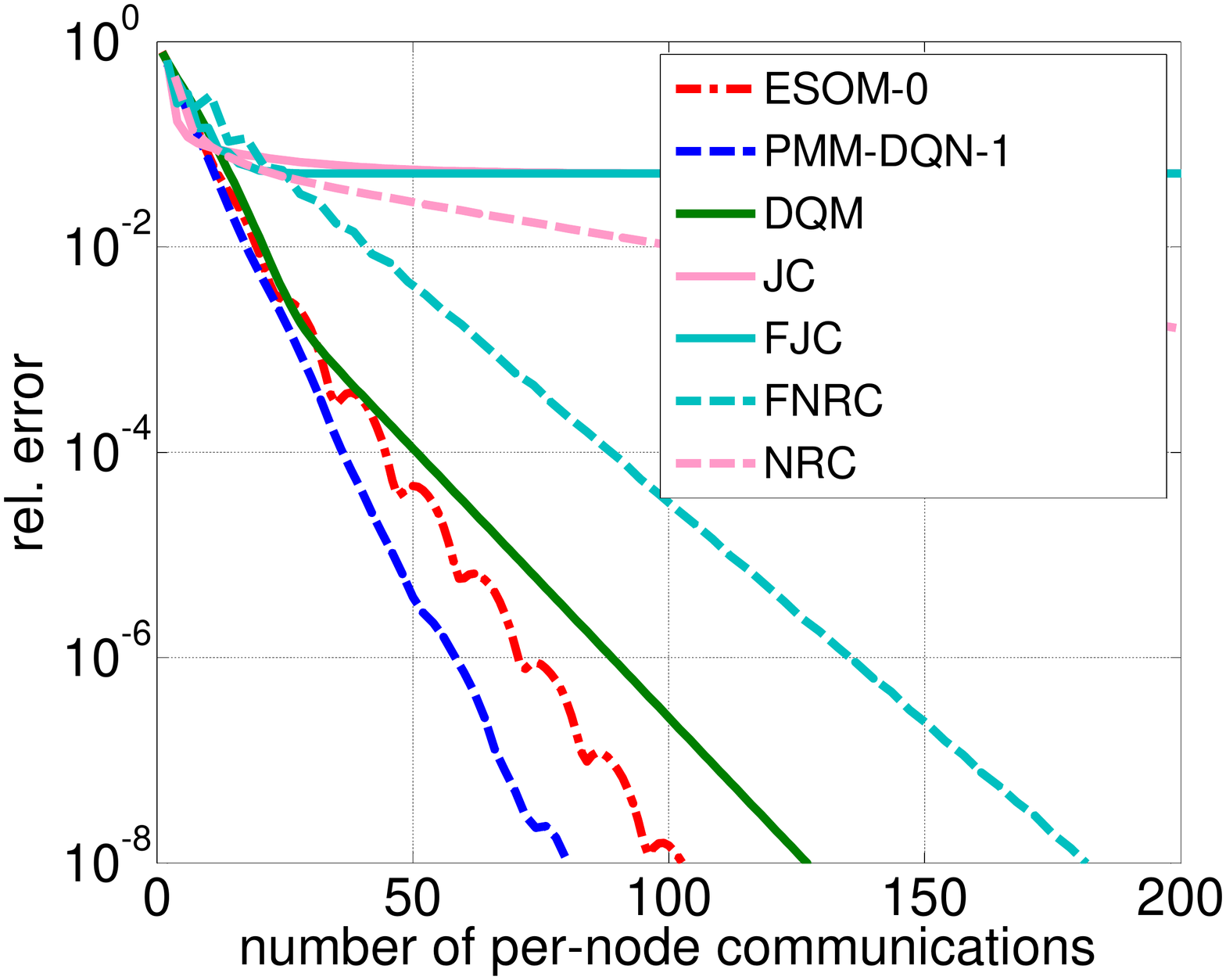}
      \caption{Comparison between the DQM
      algorithm in~\cite{DQM},
      the NRC algorithm in~\cite{NewtonRaphsonConsensus},
      the best among the
      PMM-DQN-$\ell$ algorithms,
      and the best among the ESOM-$\ell$ algorithms. The Figures plot relative error versus number of iterations~$k$ (left) and versus number of communications (right)
      for quadratic costs and $n=30$-node network.}
      \label{Figure_ESOM_2}
\end{figure}

\section{Conclusions}

The problem under consideration is defined by an aggregate, network-wide sum cost function
across nodes in a connected network.
It is assumed that the cost functions are convex and differentiable, while the network is characterized by a symmetric, stochastic matrix $ W $ that fulfils standard assumptions. The proposed methods are designed by exploiting a penalty reformulation of the original problem and rely heavily on the sparsity structure of the Hessian. The general method is tailored as a Newton-like method, taking the block diagonal part of the Hessian as an approximate Hessian and then correcting this approximation by a diagonal matrix~$\mathbb{L}_k$.
 The key point in the proposed class of methods is to exploit the structure of Hessian and replace the dense part of the inverse Hessian by an inexpensive linear approximation, determined by matrix~$\mathbb{L}_k$.
 Depending on the choice of $\mathbb{L}_k$, one can define different methods, and three of such choices are analyzed in this work. An important characteristic of the whole class of DQN methods is global linear convergence with a proper choice of the step size. Furthermore, we have shown local linear convergence for the full step size using the convergence theory of Inexact Newton methods as well as global convergence with the full step size for the special case of strictly convex quadratic loss functions.

The three specific methods are analyzed in detail, termed DQN-0, DQN-1 and DQN-2. They are defined by the three different choices of matrix~$\mathbb{L}_k$ -- the zero matrix, a constant matrix, and the iteration-varying matrix that defines a search direction which mimics the Newton direction as much as possible under the imposed restrictions of inexpensive distributed implementation. For the last choice of the time varying matrix, we have shown local linear convergence for the full step size without safeguarding.

The cost in terms of computational effort and communication of these three methods correspond to the costs of the state-of-the-art Network Newton methods, NN-0, NN-1 and NN-2, which  are used as the benchmark class in this paper. The simulation results on two relevant problems, the quadratic loss and the logistic loss, demonstrate the efficiency of the proposed methods and compare favorably with the benchmark methods.
 Finally, applying the recent contributions of~\cite{ESOM}, the proposed distributed second order methods were extended
 to the framework of proximal multiplier methods. Unlike DQN,
 the modified methods converge to the exact solution and further
 enhance the performance when high solution accuracies are required.

{\bf Acknowledgement. } We are grateful to the anonymous referees whose comments and suggestions helped us to improve the quality of this paper.

 \section*{Appendix}

Following~\cite{ESOM}, we briefly explain how
we derive the PMM-DQN methods.
%
%
%
%
 The starting point for PMM-DQN is
 the quadratically approximated
 PMM method, which takes the following form (see~\cite{ESOM}
 for details and derivations):
 \begin{eqnarray}
 \label{eqn-PMM-1}
 \widehat{x}^{\,k+1} &=& \widehat{x}^{\,k} - \widehat{\mathbb H}_k^{\,-1}
 \left( \nabla F(\widehat{x}^{\,k}) + \widehat{q}^{\,k} + \beta (\II-\ZZ)\widehat{x}^{\,k}\right)  \\
  \label{eqn-PMM-2}
 \widehat{q}^{\,k+1} &=& \widehat{q}^{\,k} + \beta (\II - \ZZ) \widehat{x}^{\,k+1}.
 \end{eqnarray}
 Here, $\beta>0$ is the (dual) step size,
 $\widehat{\mathbb H}_k$ is given in~\eqref{eqn-new-Hessian},
  and $\widehat{x}^{\,k} \in {\mathbb R}^{np}$ and
 $\widehat{q}^{\,k} \in {\mathbb R}^{np}$ are
 respectively the primal and dual variables at iteration~$k=0,1,...,$ initialized
 by~$\widehat{x}^{\,0} = \widehat{q}^{\,0} =0.$

The challenge for
distributed implementation of~\eqref{eqn-PMM-1}--\eqref{eqn-PMM-2}
 is that the inverse of $\widehat{\mathbb H}_k$
 does not respect the sparsity pattern of the network.
 The ESOM methods, proposed in~\cite{ESOM},
 approximate the inverse of $\widehat{\mathbb H}_k$
 following the NN-type approximations~\cite{ribeiro}.
 Here, we extend such possibilities
 and approximate the inverse
  of $\widehat{\mathbb H}_k$ through the DQN-type approximations.
   This is defined in \eqref{eqn-new-Hessian}--\eqref{eqn-new-splitting-G}
    and Algorithm PMM-DQN in Section~6.

As noted,
the matrix $\LL_k=0$ with PMM-DQN-0,
and it is $\LL_k=\LL_0 = \mathrm{const}$
 with PMM-DQN-1, where $\LL_0$
  is the matrix from the first
  iteration of the DQN-2 method.
  It remains to derive
  $\LL_k$ for PMM-DQN-2,
  as it is given in Section~6.
    As noted in Section~6,
    we approximate the Newton equation in~\eqref{eqn-secant-new}.
The derivation steps are the same
as with DQN-2, with the following
identification:
$\nabla^2 \Phi(x^{k})$ in (\ref{eqn-newton-equation})
is replaced
with $\widehat{\mathbb H}_k$ in~\eqref{eqn-secant-new},
and $\nabla \Phi(x^k)$
 with $\widehat{g}^{\,k}$ in~\eqref{eqn-secant-new}.
  Then, equation~\eqref{lambdak} with DQN-2
   transforms into
   the following equation with PMM-DQN-2:
   \begin{equation}
   \label{eqn-hat-u-k}
   \widehat{\mathbb L}_k\,\widehat{u}^{\,k} = - \widehat{\mathbb H}_k^{\,-1} \widehat{u}^{\,k},
   \end{equation}
   where~$\widehat{u}^{\,k}$ is given in~\eqref{eqn-hat-u-k-def}.
   Finally, it remains to
   approximate~$\widehat{\mathbb H}_k^{\,-1}$
   through a first order Taylor approximation, as follows:
   \begin{eqnarray*}
   \widehat{\mathbb H}_k^{\,-1}
   &=&
   \left[\,\,  (\beta+\epsilon_{\mathrm{pmm}}) \left( \,\II - \left( \frac{\beta}{\beta+\epsilon_{\mathrm{pmm}}}
   \ZZ - \frac{1}{\beta+\epsilon_{\mathrm{pmm}}}\nabla^2 F(\widehat{x}^{\,k})\right)\,\right)    \,\,\right]^{-1} \\
   &\approx&
   \frac{1}{\beta+\epsilon_{\mathrm{pmm}}}
   \left[\,\,  \II + \frac{\beta}{\beta+\epsilon_{\mathrm{pmm}}} \ZZ - \frac{1}{\beta+\epsilon_{\mathrm{pmm}}} \nabla^2 F(\widehat{x}^{\,k}) \,\,\right].
   \end{eqnarray*}
   The above Taylor approximation is well defined if
   the spectral radius of matrix $\left( \frac{\beta}{\beta+\epsilon_{\mathrm{pmm}}}
   \ZZ - \frac{1}{\beta+\epsilon_{\mathrm{pmm}}}\nabla^2 F(\widehat{x}^{\,k})\right)$
    is strictly less than one.
    It is easy to verify that this will be the case if the (positive)
    parameters
    $\beta$ and $\epsilon_{\mathrm{pmm}}$
     satisfy $\beta>\frac{1}{2} \max\{0,L-\epsilon_{\mathrm{pmm}}\}$.

\end{document}